\PassOptionsToPackage{hyperfootnotes=false}{hyperref}

\documentclass[11pt]{article} 

\pdfoutput=1

\usepackage{times}
\usepackage{amsmath,amsthm,amsfonts}
\usepackage{color}
\usepackage{algorithm}
\usepackage[noend]{algorithmic}
\usepackage{graphicx}
\usepackage{psfrag}
\usepackage{natbib}
\usepackage{epstopdf}
\usepackage{fullpage}
\usepackage{hyperref}
\usepackage{authblk}
\usepackage[caption=false]{subfig}
\usepackage{xcolor}
\usepackage{diagbox}
\usepackage{relsize}
\usepackage{tikz}
\usepackage{verbatim}
\usepackage{stackengine}
\usepackage[labelfont=bf]{caption}
\usepackage{soul}
\usepackage{cancel}
\usepackage{mdframed}
\usepackage[customcolors]{hf-tikz}

\hypersetup{
    pdftitle={Complex diffusion-weighted image estimation via matrix recovery under general noise models},    
    pdfauthor={Lucilio Cordero-Grande},     
    pdfkeywords={diffusion weighted imaging, Rician bias, random matrix denoising, optimal shrinkage, asymptotic risk}, 
}

\usetikzlibrary{shapes,arrows}

\tikzstyle{block} = [draw, fill=cyan!20, rectangle, rounded corners=0.5cm,
    minimum height=3em, minimum width=5em]
\tikzstyle{input} = [coordinate]

\pdfpageattr {/Group << /S /Transparency /I true /CS /DeviceRGB>>} 

\title{Complex diffusion-weighted image estimation via matrix recovery under general noise models}

\date{}
\author{L Cordero-Grande}
\author[]{D Christiaens}
\author[]{J Hutter}
\author[]{AN Price}
\author[]{JV Hajnal}
\affil[]{Centre for the Developing Brain and Biomedical Engineering Department\\School of Biomedical Engineering and Imaging Sciences\\King's College London, King's Health Partners, St Thomas' Hospital, London, SE1 7EH, UK}
\affil[]{\small{\{lucilio.cordero\_grande,daan.christiaens,jana.hutter,anthony.price,jo.hajnal\}@kcl.ac.uk}}

\definecolor{naranja}{rgb}{1.0,0.5,0}
\definecolor{violeta}{rgb}{0.5,0,1}
\definecolor{verde}{rgb}{0.75,1,0.75}
\definecolor{rojo}{rgb}{1,0.75,0.75}
\definecolor{azul}{rgb}{0.75,0.75,1}
\definecolor{naranjal}{rgb}{1.0,0.85,0.7}

\tikzstyle{nosep}=[inner sep=0pt, outer sep=0pt]
\hfsetfillcolor{naranjal}
\hfsetbordercolor{naranjal}

\begin{document}

\maketitle

\begin{abstract}

We propose a patch-based singular value shrinkage method for diffusion magnetic resonance image estimation targeted at low signal to noise ratio and accelerated acquisitions. It operates on the complex data resulting from a sensitivity encoding reconstruction, where asymptotically optimal signal recovery guarantees can be attained by modeling the noise propagation in the reconstruction and subsequently simulating or calculating the limit singular value spectrum. Simple strategies are presented to deal with phase inconsistencies and optimize patch construction. The pertinence of our contributions is quantitatively validated on synthetic data, an in vivo adult example, and challenging neonatal and fetal cohorts. Our methodology is compared with related approaches, which generally operate on magnitude-only data and use data-based noise level estimation and singular value truncation. Visual examples are provided to illustrate effectiveness in generating denoised and debiased diffusion estimates with well preserved spatial and diffusion detail. 

\end{abstract}

\begin{keywords}
diffusion weighted imaging, Rician bias, random matrix denoising, optimal shrinkage, asymptotic risk
\end{keywords}


\section{Introduction}

\label{sec:INTR}


Signal to noise ratio (SNR) is the ultimate limiting factor for high resolution magnetic resonance imaging (MRI)~\citep{Edelstein86,Fuderer88}. Diffusion weighted imaging (DWI), aiming at probing the microstructure as given by the diffusivity of water molecules, here with a focus on brain applications, is particularly SNR limited, which is due to the exponential decay of the water signal with the diffusion sensitization factor $b$. In addition, accelerated encodings have been proposed for denser diffusion sampling, which generally reduce the SNR per volume but increase the SNR per unit time~\citep{Setsompop12}. This has motivated an extensive literature on filtering methods for DWI that have evolved from standard filtering techniques on a per-volume basis~\citep{Basu06} to more comprehensive procedures that jointly consider the information sampled in the diffusion space~\citep{Tristan-Vega10}. In parallel, different spatial and diffusion representations have been suggested for DWI, which could aid denoising methods based on prior models. Recently, principal component analysis (PCA) based methods have been proposed for DWI denoising~\citep{Pai11,Manjon13}. These techniques are based on the assumption that sampling has been made redundant enough so that the covariance of the signal when arranged in a matrix of local spatial patches of $M$ elements times the number of sampled volumes $N$ can be explained by a number of components $R$ much lower than $\min(M,N)$, with the remaining observed covariance being attributable to noise. Further contributions have proposed to estimate $R$ by resorting to random matrix theory, namely to the Mar\v{c}enko-Pastur law that governs the empirical spectral noise distribution, both for DWI noise estimation~\citep{Veraart16a} and denoising~\citep{Veraart16b}. Assuming that the number of components is much lower than the matrix dimensions and the signal is strong enough, the noise will be confined into an interval of well separated small eigenvalues, so that a truncation threshold can be defined from a statistical viewpoint.

Most of the denoising techniques for DWI have been applied to magnitude-only datasets provided by scanner reconstructions, the most widely accessible image data type. However, in magnitude reconstructions the noise statistics depend on the signal level, especially at low SNR. This is typically addressed by post-processing or iterative techniques for Rician bias correction~\citep{Koay06} but these may only provide moderate benefits, particularly when considering the price of significantly increased modeling and computing complexity. Following a different path that recognizes the complex nature of the acquired data simplifies the statistical treatment, particularly in very low SNR regimes, potentially the most rewarding situation when denoising. This was demonstrated a long time ago~\citep{Bernstein89,Wood99} and applied to wavelet-based DWI denoising on a per volume basis in \cite{Wirestam06}. Additional studies have investigated the negative influence of the bias introduced by magnitude data in the reliability of derived diffusion features~\citep{Jones04,Bernd09}.

By founding the denoising criterion in universal properties of noise and being model-free, random matrix based approaches are attractive tools for denoising the complex raw data right after the reconstruction. At this stage, the statistical noise properties can be well characterized and, when conveniently arranged, the signal may admit a compact representation by singular value decomposition (SVD). In contrast, model-based approaches, unless strongly founded in diffusion data properties, may be more prone to suppress relevant information or to introduce spurious features, with magnified risks of altering the diffusion estimates considering usual needs for subsequent data pre-processing. Convincing results were provided in~\cite{Veraart16b}, for instance, about the preservation of high resolution data features after filtering.

Consequently, we propose an extension of the random matrix denoising Mar\v{c}enko-Pastur PCA (MPPCA) method in~\cite{Veraart16b}. This extension departs from homoscedastic and uncorrelated noise and hard separability of signal and noise assumptions of MPPCA by making use of the results in~\cite{Benaych-Georges12} for optimal singular value shrinkage under a general Mar\v{c}enko-Pastur law. Usage of these results allows our method to be applied in general acquisition settings not covered by the independent and identically distributed noise model, such as partial Fourier and temporally alternating encodings, and to replace the truncation operation with optimal shrinkage rules stemming from the noise model under consideration. Thus, we refer to our proposal as generalized singular value shrinkage (GSVS) denoising. Our pipeline preserves noise additivity by operating in the complex domain, with a strategy being introduced to limit the signal complexity due to unpredictable temporal phase variations. In addition, random matrix theory results are fully exploited by making use of a criterion to determine optimal patch sizes (\S~\ref{sec:THEO}). In vivo experiments are performed on high $b$-value adult data, a challenging dataset comprised of highly accelerated neonatal DWI acquired with a particularly involved encoding strategy to maximize data robustness against sporadic motion and distortions, and strongly motion degraded low SNR fetal DWI acquisitions. Benefits and/or increased flexibility of modeling the noise properties by accounting for the adopted reconstruction operations, working with complex data while correcting for linear phase inconsistencies, modelling the spectral noise properties, and using singular value shrinkage versus singular value truncation are quantitatively assessed by resorting to estimation risks provided by the random matrix theory. Moreover, we validate our approach using simulations and added value in retrieved diffusion information and derived measures is visually illustrated (\S~\ref{sec:REVA}). The source code of a \textsmaller{\textsc{MATLAB}} implementation of the signal recovery method and the data required to reproduce some of the results in Figs.~\ref{fig:NOPS},~\ref{fig:SIMU},~\ref{fig:DPCQ} and~\ref{fig:ERAD} is made available at \url{https://github.com/mriphysics/complexSVDShrinkageDWI/releases/tag/1.1.0}. Finally, we discuss the strengths and limitations of the proposed approach and provide some lines for improvement and extension (\S~\ref{sec:DISC}) to end up with some concluding remarks (\S~\ref{sec:CONC}).


\section{DWI estimation as a matrix recovery problem}

\label{sec:THEO}

DWI estimation in the presence of noise can be formulated as a set of matrix recovery subproblems by arranging the DWI data into a set of matrices $\mathbf{Y}$ of size $M\times N$. Each of these matrices is assumed to be the result of the additive corruption of an underlying (approximately) low but unknown rank $R$ matrix $\mathbf{X}$ with a random matrix with Gaussian noise entries $\mathbf{W}$, so we can write $\mathbf{Y}=\mathbf{X}+\mathbf{W}$. The objective of each subproblem is to provide an estimate of the signal matrix $\hat{\mathbf{X}}$ with some guarantees on the estimation risk induced by a given loss function $L(\mathbf{X},\hat{\mathbf{X}})$ that quantifies the error between the estimated and the actual data. In what follows we examine how the DWI data can comply with the assumptions of this model and describe the proposed image estimation techniques. First we pay attention to the nature of the complex diffusion signal (\S~\ref{sec:MSIG}), then we describe the propagation of noise in the reconstruction (\S~\ref{sec:MNOI}), later on we present the adopted algorithmic pipeline (\S~\ref{sec:TALD}), and finally we describe the procedure for matrix recovery (\S~\ref{sec:MARE})

\subsection{Low-rank DWI signal}

\label{sec:MSIG}

For the low-rank condition $R\ll\min(M,N)$ to reasonably hold, we should arrange the DWI data in a way that maximizes the redundancy in the entries of the matrices $\mathbf{Y}$. Arranging the data from local patches along the rows and the diffusion measures along the columns of $\mathbf{Y}$, simultaneously taking advantage of spatial and diffusion redundancy, has been proposed for instance in~\cite{Pai11,Manjon13,Veraart16b} as the local PCA model for diffusion denoising. We can define a set of local patch extraction matrices indexed by their center voxel index $q$, $\mathbf{P}^{[q]}$, $q\in\mathcal{Q}=\{1,\ldots, Q\}$ of size $M\times Q$, with $Q$ the number of voxels in each data volume, $\mathcal{Q}$ the set of voxel indexes, and $M$ the number of voxels in the local 3D patch, which are indexed by $m\in\mathcal{M}=\{1,\ldots, M\}$. We use spherical instead of more common cubic patches, with potential benefits from rotational invariance. Thus, we make $p^{[q_1]}_{m\leftarrow q_2}=1$ if the voxel indexed by $q_2$ is the $m$-closest to the voxel indexed by $q_1$, and $0$ otherwise.

A potential source of spurious complexity in the data is given by artifacts. For complex DWI data, we have to pay special attention to the phase of the signal. The DWI phase varies in a largely unpredictable manner during the acquisition mainly due to patient motion during the scan~\citep{Bammer10}. Larger gradient strengths imply increased levels of phase fluctuations and reduced SNR, so accurate phase corrections become especially difficult for high $b$-values. Thus, our method performs phase corrections on the basis of a robust gross approximation of the underlying phase (see \S~\ref{sec:TALD}), which can increase the redundancy of the signal without introducing undesired biases. For now, we denote the reconstructed volumes for the set of diffusion measurements as $\mathbf{y}$, their phase corrected versions as $\tilde{\mathbf{y}}$, the local information for the patch centered at $[q]$ after phase correction as $\tilde{\mathbf{y}}^{[q]}=\mathbf{P}^{[q]}\tilde{\mathbf{y}}$ and express patch-to-matrix assignments as $\mathbf{Y}^{[q]}\leftarrow\tilde{\mathbf{y}}^{[q]}$. In this setting, DWI denoising involves solving $Q'\leq Q$ matrix denoising subproblems that provide a set of estimates for the signal at $q$, with one estimate for each patch the voxel $q$ belongs to, so that a patch combination rule can be implemented to build the resulting denoised DWI volumes (see \S~\ref{sec:MARE}).

\subsection{Additive Gaussian noise in DWI reconstruction}

\label{sec:MNOI}

DWI measures are assumed to follow a circularly symmetric complex Gaussian stationary noise distribution of zero mean and a given variance $\sigma^2$, $\mathcal{CN}(0,\sigma^2)$ modeling the Johnson-Nyquist noise observed in the quadrature receiver. In parallel imaging, levels and correlation of noise in each of the receivers are described by a noise covariance matrix $\boldsymbol{\Lambda}_{\mathbf{z}}$, which reads as the covariance $\boldsymbol{\Lambda}$ of the measured data $\mathbf{z}$, so the noise follows $\boldsymbol{\mathcal{CN}}(\boldsymbol{\mu}_{\mathbf{z}},\boldsymbol{\Lambda}_{\mathbf{z}})$ where $\boldsymbol{\mu}_{\mathbf{z}}=\mathbf{0}$. Image reconstruction of each of the DWI volumes $n\in\mathcal{N}=\{1,\ldots,N\}$ can be performed independently by linear inversion ---a.k.a. sensitivity encoding (SENSE) method~\citep{Pruessmann99}---:
\begin{equation}
\label{ec:SENS}
\mathbf{y}^{\text{SENSE}}_n=(\mathbf{E}_n^H\boldsymbol{\Lambda}_{\mathbf{z}}^{-1}\mathbf{E}_n)^{-1}\mathbf{E}_n^H\boldsymbol{\Lambda}_{\mathbf{z}}^{-1}\mathbf{z}_n
\end{equation}
with $\mathbf{E}_n$ the encoding operator.

Assuming for now that the encoding is the same for all volumes, the basic encoding operator in parallel MRI can be expressed as $\mathbf{E}=\mathbf{F}\mathbf{S}$, with $\mathbf{F}$ accounting for Fourier encoding (encompassing the applied spectral sampling structure) and $\mathbf{S}$ for sensitivity encoding. A whitening transformation $\boldsymbol{\Upsilon}_{\mathbf{z}}$ of the measured receiver noise satisfying $\boldsymbol{\Upsilon}_{\mathbf{z}}^H\boldsymbol{\Upsilon}_{\mathbf{z}}=\boldsymbol{\Lambda}_{\mathbf{z}}^{-1}$ is used in practice to transform the data $\overline{\mathbf{z}}=\boldsymbol{\Upsilon}_{\mathbf{z}}\mathbf{z}$ and the encoding operator $\overline{\mathbf{E}}=\boldsymbol{\Upsilon}_{\mathbf{z}}\mathbf{E}=\mathbf{F}\boldsymbol{\Upsilon}_{\mathbf{z}}\mathbf{S}=\mathbf{F}\overline{\mathbf{S}}$. This generates an equivalent inverse problem where the channel noise is decorrelated and standardized~\citep{Pruessmann01}, so we can write $\boldsymbol{\Lambda}_{\overline{\mathbf{z}}}=\boldsymbol{\mathcal{CN}}(\mathbf{0}_{KLN},\mathbf{I}_{KLN})$, with $\mathbf{0}_{KLN}$ and $\mathbf{I}_{KLN}$ denoting respectively the null vector and identity matrix of size $KLN$, where $K$ is the number of spectral samples and $L$ is the number of coils.

Due to reconstruction linearity, $\mathbf{y}$ follows $\boldsymbol{\mathcal{CN}}(\boldsymbol{\mu}_{\mathbf{y}},\boldsymbol{\Lambda}_{\mathbf{y}})$. The expected value of noise in the reconstructed data is simply
\begin{equation}
\label{ec:NPRM}
\boldsymbol{\mu}^{\text{SENSE}}_{\mathbf{y}}=(\overline{\mathbf{E}}^H\overline{\mathbf{E}})^{-1}\overline{\mathbf{E}}^H\mathbf{0}_{KL}\otimes\mathbf{1}_N=
\mathbf{0}_Q\otimes\mathbf{1}_N=\mathbf{0}_{QN},
\end{equation}
with $\otimes$ denoting the Kronecker product and $\mathbf{1}_{N}$ the all-ones vector of size $N$. On the other hand, the noise covariance is the same for all volumes:
\begin{equation}
\label{ec:NPRC}
\boldsymbol{\Lambda}^{\text{SENSE}}_{\mathbf{y}}=(\overline{\mathbf{E}}^H\overline{\mathbf{E}})^{-1}\overline{\mathbf{E}}^H
\mathbf{I}_{KL}\overline{\mathbf{E}}(\overline{\mathbf{E}}^H\overline{\mathbf{E}})^{-1}\otimes\mathbf{I}_N=
(\overline{\mathbf{E}}^H\overline{\mathbf{E}})^{-1}\otimes\mathbf{I}_N=\boldsymbol{\Sigma}_Q\otimes\mathbf{I}_N,
\end{equation}
with $\boldsymbol{\Sigma}_Q$ referring to a given covariance matrix of size $Q$.
As for patch-based denoising, we are interested on the local patch marginalization of the noise covariance:
\begin{equation}
\label{ec:NCLP}
\boldsymbol{\Lambda}_{\mathbf{Y}^{[q]}}=\mathbf{P}^{[q]}\boldsymbol{\Lambda}_{\mathbf{y}}\mathbf{P}^{[q]H}=\mathbf{P}^{[q]}\boldsymbol{\Sigma}_Q\mathbf{P}^{[q]H}\otimes\mathbf{I}_N=\boldsymbol{\Sigma}_M^{[q]}\otimes\mathbf{I}_N.
\end{equation}

Before building the random matrix recovery model, let's review the properties of noise in different reconstruction scenarios:
\begin{itemize}
	\item \textbf{Additive noise}. This property will generally hold for complex reconstructed data, as the application of a linear reconstruction operator preserves noise additivity~\eqref{ec:NPRM}. However, this additive property is broken for magnitude-only data due to the non-linearity of the magnitude operation, particularly at low SNR. Using magnitude data, the noise standard deviation will depend on the signal level in agreement to a Rician distribution~\citep{Gudbjartsson95}, which prevents a straightforward formulation of the matrix denoising problem. In a somewhat naive manner, one could still apply the singular value shrinkage to magnitude data but, as we will illustrate in \S~\ref{sec:EDDM}, for low SNR the noise bias would blend into the signal components so, even if there is some noise reduction, the diffusion contrast appears hampered.
	\item \textbf{Full noise independence}. The simplest situation occurs when the Fourier encoding operator is given by the discrete Fourier transform (DFT), $\mathbf{F}=\boldsymbol{\mathcal{F}}$, which corresponds to Cartesian sampling without acceleration and with matched acquired and reconstructed grids (identified hereafter as case FULL). Due to unitarity of the DFT operator, $\boldsymbol{\mathcal{F}}^H=\boldsymbol{\mathcal{F}}^{-1}$, the spatial noise covariance can be written as
\begin{equation}
\label{ec:YFIN}
\boldsymbol{\Lambda}^{\text{FULL}}_{\mathbf{y}}=(\overline{\mathbf{S}}^H\overline{\mathbf{S}})^{-1}\otimes\mathbf{I}_N=\boldsymbol{\Sigma}^{\text{FULL}}_Q\otimes\mathbf{I}_N.
\end{equation}
$\overline{\mathbf{S}}$ is a spatially diagonal operator, so it acts on each sampled voxel $q$ separately. Then, it can be arranged for each voxel as a column vector of size $C\times 1$, $\overline{\mathbf{S}}_{q}=\overline{s}_{c}$, and $\boldsymbol{\Sigma}^{\text{FULL}}_Q$ reduces to a diagonal matrix comprised of scalar variances:
\begin{equation}
\label{ec:LFIN}
\boldsymbol{\Sigma}^{\text{FULL}}_Q\equiv\boldsymbol{\Sigma}^{\text{FULL},\mathcal{Q}}_1=\sigma_q^2=(\overline{\mathbf{S}}^H_q\overline{\mathbf{S}}_q)^{-1},
\end{equation}
with $\sigma_q^2$ referring to the elements in the diagonal.
	\item \textbf{Local independence}. A common reconstruction setting is that of parallel imaging acceleration via homogeneous spectral undersampling (case UNDER). In this case the Fourier encoding operator can be expressed as $\mathbf{F}=\boldsymbol{\mathcal{F}}\mathbf{U}$, with $\mathbf{U}$ a spatial overlapping operator from the reconstruction to the acquisition field of view (FOV). Accordingly, the noise covariance is given by:
%
\begin{equation}
\label{ec:YPIN}
\boldsymbol{\Lambda}^{\text{UNDER}}_{\mathbf{y}}=(\overline{\mathbf{S}}^H\mathbf{U}^H\mathbf{U}\overline{\mathbf{S}})^{-1}\otimes\mathbf{I}_N=
\boldsymbol{\Sigma}^{\text{UNDER}}_Q\otimes\mathbf{I}_N.
\end{equation}
The spatial noise correlation can be represented by a sparse matrix with non-zero entries only along the main diagonal and the $U-1$ element pairs corresponding to $U$ aliased voxels in the reduced acquisition FOV. Considering a given aliased dimension $i$, these correlations will only affect those voxels separated by $uK_i$, with $u\in\mathcal{U}=\{0,\ldots,U-1\}$ and $K_i$ the number of acquired samples along dimension $i$. One can define a sensitivity matrix\footnote{This matrix is assumed to encompass the normalization from the acceleration ratio between accelerated and non-accelerated acquisitions, which is defined as the number of samples of a full acquisition divided by the number of samples of the accelerated scan.} $\overline{\mathbf{S}}_{\mathbf{q}_{\mathcal{U}}}=\overline{s}_{c,u}$ for the overlapped voxels $\mathbf{q}_{\mathcal{U}}=\{q_0,\ldots q_{U-1}\}$ so that noise covariances are given by
\begin{equation}
\label{ec:LPIN}
\boldsymbol{\Sigma}^{\text{UNDER}}_Q\equiv\boldsymbol{\Sigma}^{\text{UNDER},\mathcal{Q}/\mathcal{U}}_{U}=\sigma^2_{q_{u_1},q_{u_2}}=(\overline{\mathbf{S}}^H_{\mathbf{q}_{\mathcal{U}}}\overline{\mathbf{S}}_{\mathbf{q}_{\mathcal{U}}})^{-1},
\end{equation}
with $\mathcal{Q}/\mathcal{U}$ the quotient of $\mathcal{Q}$ by $\mathcal{U}$. The local covariance matrix in~\eqref{ec:NCLP} becomes diagonal due to local independence, and we can write $\boldsymbol{\Sigma}^{\text{UNDER},\mathcal{Q}}_1=\sigma^2_{q_{\mathcal{U}},q_{\mathcal{U}}}$, computed using the set of overlapped voxels that includes the voxel $q$. Normalization by this matrix has been previously proposed as the reconstruction in SNR units~\citep{Kellman05} and, for accelerated acquisitions, the ratio between $\sigma_{q_{\mathcal{U}},q_{\mathcal{U}}}$ and $\sigma_q$ corresponds to the g-factor times the square root of the sampling ratio at voxel $q$~\citep{Pruessmann99}.
	\item \textbf{Spatial correlations}. More general settings arise for more complex reconstruction techniques or when the data is acquired in a non uniform manner. Typical examples include when the Fourier operator involves regridding, or Gibbs ringing filtering and/or zero-filling interpolation are applied to the reconstructed data. These situations will generally break the noise independence. The specific noise properties will depend on the reconstruction operator. However, it is generally possible to completely characterize the noise propagation through a linear reconstruction method by making use of a given noise covariance matrix over the local patches, as defined in~\eqref{ec:NCLP}, which can be obtained either analytically or by Monte Carlo simulation. Here, as an example of interest in our data, we focus on partial Fourier (PF) or half scan acquisitions (case HALF) where the reconstruction model in~\eqref{ec:SENS} is extended by adding a spectral filter $\mathbf{G}$, for instance the ramp filter used for homodyne detection in~\cite{Noll91} or simply a zero-filling filter. While standard PF reconstruction involves a separate phase estimation step, this can be integrated in the signal retrieval formulation within our setting, so we only need to include the k-space weighting required to retrieve the target signal resolution. The noise covariance becomes:
\begin{equation}
\label{ec:YHIN}
\boldsymbol{\Lambda}^{\text{HALF}}_{\mathbf{y}}=\mathbf{G}(\overline{\mathbf{S}}^H\mathbf{U}^H\mathbf{U}\overline{\mathbf{S}})^{-1}\mathbf{G}^H\otimes\mathbf{I}_N=
\boldsymbol{\Sigma}_Q^{\text{HALF}}\otimes\mathbf{I}_N.
\end{equation}
	\item \textbf{Temporal heteroscedasticity}. This can be the case when the encoding structure is non stationary, typically from an acquisition technique where more evenly distributed net information is pursued by alternating the encodings for different excitations. Here we focus on the method we have used in our neonatal cohort~\citep{Hutter18}, where the phase encoding (PE) direction is interleaved volumewise among the four possible Cartesian in-plane directions for increased tolerance to main field inhomogeneity induced distortions~\citep{Brushan14}. We can write the reconstruction for this case (case INTER) using the most general noise model of those previously described, HALF, as
	\begin{equation}
	\label{ec:STIX}
	\mathbf{y}^{\text{INTER}}_n=\mathbf{G}_{a(n)}(\overline{\mathbf{S}}_{a(n)}^H\mathbf{U}_{a(n)}^H\mathbf{U}_{a(n)}\overline{\mathbf{S}}_{a(n)})^{-1}\overline{\mathbf{S}}_{a(n)}^H\mathbf{U}_{a(n)}^H\mathbf{F}^H\overline{\mathbf{z}}_n,
	\end{equation}
	where $a(n)$ indexes the encoding used for volume $n$, with $a\in\mathcal{A}=\{1,\ldots,A\}$. In this situation we can arrange the covariance between the matrix entries as a fourth order tensor:
	\begin{equation}
	\label{ec:YMIN}
	\boldsymbol{\Lambda}^{\text{INTER}}_{\mathbf{y}}=\sum_{a\in\mathcal{A}}\boldsymbol{\Sigma}^{\text{HALF},a}_Q\otimes\boldsymbol{\Sigma}^a_N,
	\end{equation}
	where $\boldsymbol{\Sigma}^{\text{HALF},a}_Q$ represents the spatial covariance matrix for the encoding $a$ and $\boldsymbol{\Sigma}_N^a$ is an indicator diagonal matrix with ones in those entries that correspond to the volumes acquired with encoding $a$.
\end{itemize}

\subsection{Denoising algorithm}

\label{sec:TALD}

Figure~\ref{fig:DAST} sketches the steps of our denoising procedure. Here is a detailed description:
\begin{figure}[!htb]
\begin{center}\resizebox{1.08\textwidth}{!}{\hspace{1cm}
\input{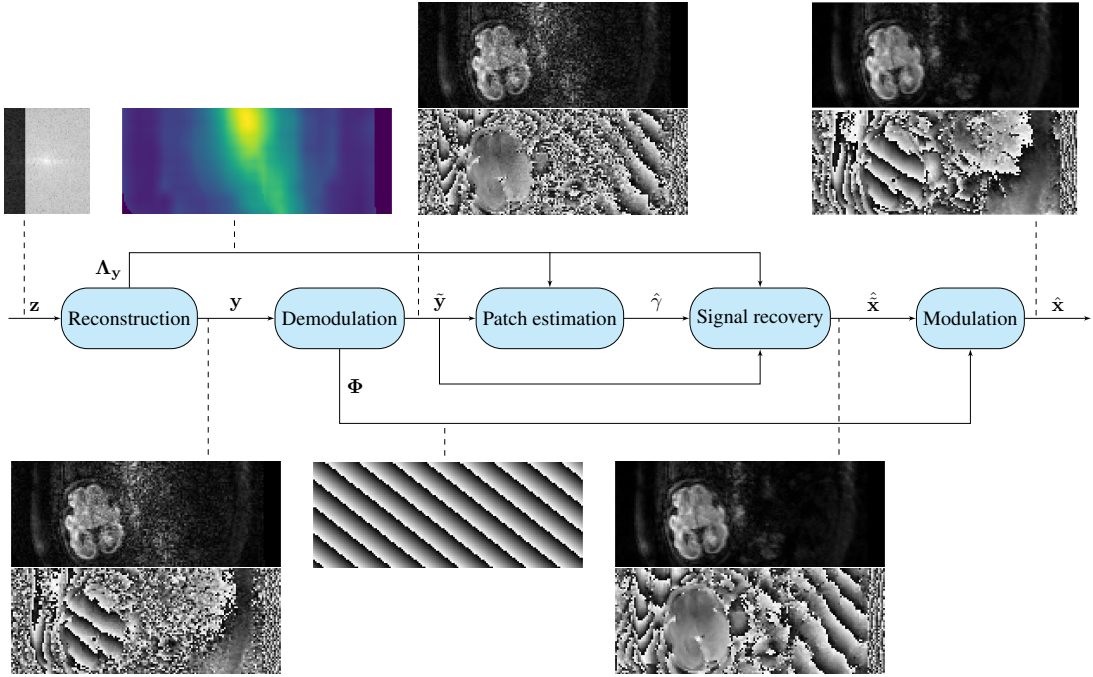}}\end{center}
\caption{Block diagram of the DWI denoising algorithm exemplified with a fetal dataset. Note PF data $\mathbf{z}$ has been acquired which produces a non-symmetric spectrum in the PE (horizontal) dimension. After the inversion of the DFT and spatial unfolding during the reconstruction, we have access to the magnitude and phase of $\mathbf{y}$ and the noise covariance $\boldsymbol{\Lambda}_{\mathbf{y}}$, here illustrated by the spatial noise amplification levels. Later on, a linear phase $\boldsymbol{\Phi}$ is estimated and removed from the data, so we obtain $\tilde{\mathbf{y}}$. Optimal patch size $\hat{\gamma}$ is estimated to perform signal prediction, $\hat{\tilde{\mathbf{x}}}$, and phase demodulation is reversed to provide the final estimate $\hat{\mathbf{x}}$.}
\label{fig:DAST}
\end{figure}
\begin{itemize}
	\item \textbf{Reconstruction}. We use the SENSE reconstruction framework in~\eqref{ec:SENS}. Our datasets contain both in-plane and simultaneous multi-slice (SMS) accelerated echo planar imaging acquisitions, so the reconstruction method follows the extended SENSE technique in~\cite{Zhu16}. Noise correlation introduced by the readout gradient ramps is considered negligible. Sensitivity estimates from a conventional reference scan are refined with the information from non-SMS reference acquisitions with matched readouts~\citep{Hennel16} to promote matched coil map and image distortions, which explains why the sensitivity matrices $\overline{\mathbf{S}}$ depend on the applied encoding in~\eqref{ec:STIX}. Sensitivity estimation uses the variational formulation in~\cite{Allison13}.
	\item \textbf{Demodulation}. A model for phase corruption of the DWI signal in the presence of bulk motion was derived in~\cite{Anderson94}. Using this model, partial correction of the temporal phase variations can be achieved by estimating a linear phase $\boldsymbol{\Phi}$ for each of the reconstructed slices, and subtracting it from the data, $\tilde{\mathbf{y}}=\boldsymbol{\Phi}^H\mathbf{y}$. By assuming stationarity and employing the Fourier shift theorem, the linear phase ramp and offset are respectively estimated as the harmonic corresponding to the position of the maximum of the Fourier transform of the complex image and the phase at this maximum. This simple phase model is expected to be robust enough to be applied in low SNR regimes and to largely reduce the dynamic phase fluctuations, thus increasing the redundancy of the data and consequently reducing the rank of the matrices to be recovered. Phase correction of linear ramps in the PE direction may alter the temporal noise properties for PF, which could be incorporated into a model like~\eqref{ec:YMIN}, but for computational simplicity we assume that this effect is negligible.	
	\item \textbf{Patch estimation}. The choice of the patch dimension $M$, or equivalently the matrix aspect ratio $\gamma=M/N$, may impact recovery performance. We propose to optimize the patch size by minimizing the relative asymptotic mean squared error (RAMSE) of matrix estimation over a set of candidate $\gamma_{(j)}: 0<\gamma_{(j)}<\infty, j\in\mathcal{J}=\{1,\ldots,J\}$. The procedure is analogous to matrix recovery, so both will be described in~\S~\ref{sec:MARE}.
	\item \textbf{Signal recovery}. The optimal aspect ratio $\hat{\gamma}$ is used to build the patches for matrix recovery, which allows a denoised estimate $\hat{\tilde{\mathbf{x}}}$ to be obtained from the observed data $\tilde{\mathbf{y}}$.
	\item \textbf{Modulation}. The phase subtraction performed in the demodulation step can be reversed to obtain an estimate of the complex DWI signal, $\hat{\mathbf{x}}=\boldsymbol{\Phi}\hat{\tilde{\mathbf{x}}}$.
\end{itemize}

\subsection{Matrix recovery}

\label{sec:MARE}

In this Section we describe both the procedure for matrix-based signal recovery and the related patch size estimation technique. The generic recovery pipeline is sketched in Fig.~\ref{fig:DSRP}. Its inputs are the reconstructed and demodulated data from one of the reconstruction techniques described in~\S~\ref{sec:MNOI}, $\tilde{\mathbf{y}}$, the noise covariance induced by the reconstruction, $\boldsymbol{\Lambda}_{\mathbf{y}}$, and the optimal aspect ratio provided by the patch size estimation procedure described at the end of this Section, $\hat{\gamma}$. Here we summarize its blocks:
\begin{figure}[!htb]
\resizebox{0.95\textwidth}{!}{\input{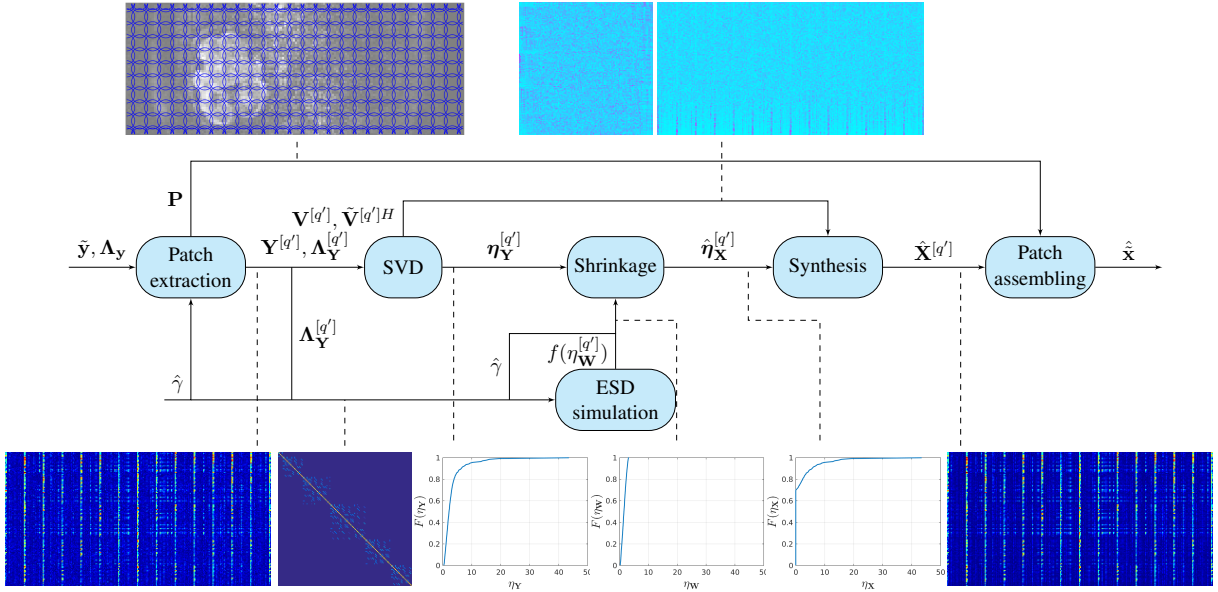}}
\caption{Block diagram of the signal recovery procedure. Local information within ball-shaped sliding patches is extracted by $\mathbf{P}$. Random signal plus noise matrices $\mathbf{Y}$ and covariance tensors $\boldsymbol{\Lambda}_{\mathbf{Y}}$ are constructed for each of the patches $[q']$. SVD of these matrices is performed with singular vectors in $\mathbf{V}$ and $\tilde{\mathbf{V}}^H$ and singular values in $\boldsymbol{\eta}_{\mathbf{Y}}$. In parallel, noise properties are modelled by a spectral distribution $f(\boldsymbol{\eta}_{\mathbf{W}})$, which allows optimally shrunk singular values, $\hat{\boldsymbol{\eta}}_{\mathbf{X}}$, to be computed. Matrix estimates $\hat{\mathbf{X}}$ are synthesized using the shrunk singular values and overlapped patch information is assembled onto the signal estimates $\hat{\tilde{\mathbf{x}}}$.}
\label{fig:DSRP}
\end{figure}
\begin{itemize}
	\item \textbf{Patch extraction}. Patch extraction uses the procedure described in \S~\ref{sec:MSIG}. To accelerate computation, the number of patches to cover the FOV is lower than the total number of voxels, $Q'\leq Q$. In our default implementation, patch sliding is performed with a subsampling factor $T=2$ in all directions, so $Q'\simeq Q/T^3$, with negligible impact in quality. This step outputs a patch extraction operator $\mathbf{P}$, used later for patch assembling, the matrices to be recovered, with local information for a given patch arranged along the rows and corresponding diffusion measures along the columns, $\mathbf{\mathbf{Y}}^{[q']}$, and a set of local noise covariance matrices, $\boldsymbol{\Lambda}_{\mathbf{Y}}^{[q']}$.
	\item \textbf{SVD}. SVDs are computed for each patch matrix $[q']$, $\mathbf{Y}^{[q']}=\mathbf{V}^{[q']}\boldsymbol{\Theta}^{[q']}\tilde{\mathbf{V}}^{[q']H}$, with $\mathbf{V}^{[q']}$ and $\tilde{\mathbf{V}}^{[q']H}$ containing respectively the left and right singular vectors and $\boldsymbol{\Theta}^{[q']}$ a diagonal matrix with the singular values $\boldsymbol{\eta}_{\mathbf{Y}}^{[q']}$ in its diagonal.
	\item \textbf{Shrinkage}. The matrix recovery problem via SVD is grounded on the related spiked covariance model~\citep{Johnstone01}. This model was proposed to study the distribution of eigenvalues of sample covariance matrices\footnote{Without lack of generality we present the model for $M/N=\gamma<1$.} $N^{-1}\mathbf{Y}\mathbf{Y}^H$ drawn from population covariances whose eigenvalues $\boldsymbol{\eta}^2_{\mathbf{X}}$ are all equal to $1$ rather than for a small subset of them $\boldsymbol{\eta}^2_{\mathbf{X}r}$, $1\leq r \leq R$, $R\ll M$, that satisfy $\eta^2_{\mathbf{X}1}\geq\ldots\geq\eta^2_{\mathbf{X}R}>1$ and are independent of the number of variables $M$ and observations $N$. Then, defining an asymptotically constant ratio of variables and observations $\gamma=M/N$, it has been shown~\citep{Baik06} that, whenever $\eta_{\mathbf{X}r}^2>1+\sqrt{\gamma}$, the spiky eigenvalues of the sample covariance matrix $\eta^2_{\mathbf{Y}r}$ tend to a deterministic limit when $N\rightarrow\infty$ given by $\eta^2_{\mathbf{Y}r}=\eta^2_{\mathbf{X}r}(1+\gamma/(\eta^2_{\mathbf{X}r}-1))$; i.e., there is a positive bias in the sample spiky eigenvalues as compared to the eigenvalues of the population $\eta^2_{\mathbf{X}r}$. Building on this model, a set of optimal shrinkage criteria $\hat{\eta}^2_{\mathbf{X}r}=h(\eta^2_{\mathbf{Y}r})$ for different loss functions $L(\hat{\mathbf{X}},\mathbf{X})$ have been derived in~\cite{Donoho18}. In this paper we focus on the Frobenius loss criterion
	\begin{equation}
	\label{ec:FRLO}
	L_{\text{FROB}}(\hat{\mathbf{X}},\mathbf{X})=\|\hat{\mathbf{X}}-\mathbf{X}\|_2^2
	\end{equation}
	but keeping in mind that similar ways of proceeding are possible for other losses.
	
	The condition of almost all population eigenvalues equal to $1$ in~\cite{Donoho18} makes this covariance estimation problem analogous to the matrix denoising problem in the presence of additive independent and identically distributed (i.i.d.) Gaussian noise, for which optimal shrinkers for different cost functions have been derived in~\cite{Gavish17}. However, the description of noise propagation in the reconstruction in~\S~\ref{sec:MNOI} shows that the i.i.d. condition is generally not valid in DWI reconstructed data.~\cite{Benaych-Georges12}, using a more general model than~\cite{Gavish17}, studied the asymptotic limits of the singular values of additive perturbations of zero-mean Gaussian not necessarily i.i.d. noise-only matrices $\mathbf{W}$. They showed that, assuming a perturbation signal matrix drawn from a bi-unitarily invariant distribution, these limits only depend on integral transforms of the empirical spectral distribution (ESD) of noise, that is to say, on the distribution of the singular values of matrices with noise-only entries.~\cite{Nadakuditi14} proposed to use those limits to build optimal shrinkers for the observed singular values when using unitarily invariant costs such as~\eqref{ec:FRLO}.
	
	Following~\cite{Benaych-Georges12}, we can define the D-transform as
	\begin{equation}
	\label{ec:DTRA}
	D_f(x)=\left(\int_{-\infty}^{\infty}\frac{x}{x^2-\eta_{\mathbf{W}}^2}f(\eta_{\mathbf{W}})d\eta_{\mathbf{W}}\right)\left(\gamma\int_{-\infty}^{\infty}\frac{x}{x^2-\eta_{\mathbf{W}}^2}f(\eta_{\mathbf{W}})d\eta_{\mathbf{W}}+\frac{1-\gamma}{x}\right), x>\eta_{\mathbf{W}}^{+}.
	\end{equation}
	with the ESD probability density $f(\eta_{\mathbf{W}})$ ---or alternatively the ESD cumulative density $F(\eta_{\mathbf{W}})$--- depending on the noise covariance matrix and the aspect ratio as described in the next item, and $\eta_{\mathbf{W}}^{+}$ referring to the upper bound of the support of $f(\eta_{\mathbf{W}})$. Then, it turns that from the set of all possible shrinkers, the one that asymptotically minimizes the expected Frobenius loss in~\eqref{ec:FRLO} can be obtained in terms of the ratio of the D-transform of the observed singular values above the detection threshold and its derivative~\citep{Nadakuditi14}:
	\begin{equation}
	\label{ec:OSGE}
	\hat{\eta}_{\mathbf{X}r}=\begin{cases}\displaystyle -2\frac{D_f(\eta_{\mathbf{Y}r})}{D'_f(\eta_{\mathbf{Y}r})}& \mbox{if }\eta_{\mathbf{Y}r}>\eta_{\mathbf{W}}^{+}\\0 & \mbox{otherwise,}\end{cases}
	\end{equation}	
	Note that defining the set of non-fully suppressed components $\mathcal{R}=\{r:\eta_{\mathbf{Y}r}>\eta_{\mathbf{W}}^{+}\}$, a rank estimate is obtained as $\hat{R}=|\mathcal{R}|$, with $|\cdot|$ denoting the set cardinality. Consequently, in contrast to other DWI denoising techniques~\citep{Pai11,Manjon13,Veraart16b}, our proposal does not impose either truncation or full preservation of the SVD components, as truncation thresholds, even when optimally chosen, have been shown to compare unfavourably with optimal shrinkage~\citep{Gavish14,Gavish17}. The benefit of shrinking with respect to truncating is relevant in situations where a given component may include comparable contributions of signal and noise, which we have seen to generally occur in our datasets.
	\item \textbf{ESD modelling}. The Mar\v{c}enko-Pastur theorem~\citep{Marcenko67,Silverstein95} states that the ESD of the population covariance matrix $\boldsymbol{\Sigma}_M$ converges almost surely to a bounded distribution that can be described by a fixed point equation. Numerical tools~\citep{Dobriban15,Ledoit17} have been developed to compute the ESD for noise covariance matrices of the form $\boldsymbol{\Lambda}_{\mathbf{Y}}=\boldsymbol{\Sigma}_{M}\otimes\mathbf{I}_N$, which applies to the stationary encoding models in~\eqref{ec:LFIN},~\eqref{ec:LPIN}~and~\eqref{ec:YHIN}. Although analogous fixed point equations~\citep{Wagner12} and eigenvalue confinement guarantees~\citep{Kammoun16} are available for more general noise models such as those of non stationary encodings in~\eqref{ec:YMIN}, we do not know of any numerical tool developed for general random interactions between the matrix entries. Thereby, here we adopt a more flexible approach that approximates the asymptotic spectrum of noise by Monte Carlo simulations~\citep{Jing10}. Namely, we draw $Q'$ matrices $\mathbf{W}$ of size $B(M\times N)$, with $B$ a positive integer factor to guarantee that the spectrum of these matrices is a good enough approximation to their asymptotic limit. If these matrices are synthesized so that they follow the noise distribution in the data, $\boldsymbol{\mathcal{CN}}(\boldsymbol{0},\boldsymbol{\Lambda}_{\mathbf{Y}^{[q']}})$, we can compute their decreasingly sorted singular values $\hat{\boldsymbol{\eta}}_{\mathbf{W}}^{[q']}$ and make the following approximation for each patch $[q']$:
	\begin{equation}
	\label{ec:AESD}
	f(\eta_{\mathbf{W}})\simeq\frac{1}{BM}\sum_{m=1}^{BM}\delta(\eta_{\mathbf{W}}-\hat{\eta}_{\mathbf{W}m}),
	\end{equation}
	with $\eta_{\mathbf{W}}^{+}\simeq\hat{\eta}_{\mathbf{W}1}$ and $\delta$ the Dirac delta. Note that, as a result of our investigations to directly compute the asymptotic noise spectrum, we have also developed some techniques for solving the generalized Mar\v{c}enko-Pastur equation in the setting of~\eqref{ec:YMIN}. Although due to computational inefficiency these developments have not been adopted for our problem, we have collected them in a companion note~\citep{Cordero-Grande18}.
	\item \textbf{Synthesis}. Estimated signal matrices are reconstructed for each patch using the shrunk singular values, $\hat{\mathbf{X}}^{[q']}=\mathbf{V}^{[q']}\hat{\boldsymbol{\Theta}}^{[q']}\tilde{\mathbf{V}}^{[q']H}$, with $\hat{\boldsymbol{\Theta}}^{[q']}$ a diagonal matrix built from the singular values $\hat{\boldsymbol{\eta}}_{\mathbf{X}}^{[q']}$ obtained by~\eqref{ec:OSGE}.
	\item \textbf{Patch assembling}. The estimated DWI volumes are obtained as 
	\begin{equation}
	\label{ec:SSPA}
	\hat{\tilde{\mathbf{x}}}=\hat{\tilde{x}}_{q_2,n}=\displaystyle\sum_{q_1\in\mathcal{Q}'}\omega_{q_2}^{[q_1]}\hat{X}_{m\rightarrow q_2,n}^{[q_1]},
	\end{equation}
	with $\omega_{q_2}^{[q_1]}$ satisfying $\displaystyle\sum_{q_1'\in\mathcal{Q}'}\omega_{q_2}^{[q_1']}=1$ used to weight the matrix estimate from the patch centered at $q_1$ in the $q_2$ assembled estimate. This weighting can be chosen to be uniform, depending on the inverse of the variance of the estimates, as in~\cite{Dabov07}, which in our setting could be obtained from~\eqref{ec:AMGE} below, or based on the distance between $q_2$ and $q_1$, for instance by means of a Gaussian window. Simulation tests in~\S~\ref{sec:SIMU} were run using inverse variance combination, which provided the best results. However, the differences with both uniform and Gaussian kernel combination were small, so in the in vivo tests in~\S\S~\ref{sec:PHST},~\ref{sec:GENO}, and \ref{sec:EDDM} we opted for a Gaussian window combination, which was judged potentially beneficial in the presence of motion.
\end{itemize}

The procedure for patch size estimation is sketched in Fig.~\ref{fig:DPEP}. It uses the same inputs and analogous methods than the decomposition for matrix recovery, but the objective is to select the patch size that yields the best RAMSE. In what follows we describe the implementation details and the formulas for RAMSE estimation:
\begin{figure}[!htb]
\begin{center}\resizebox{0.6\textwidth}{!}{
\input{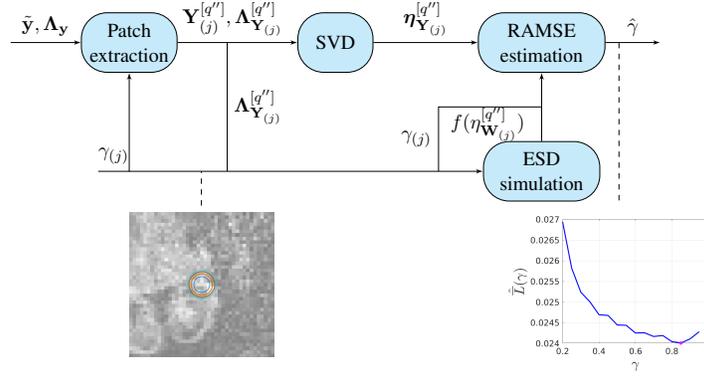}}\end{center}
\caption{Block diagram of the patch estimation procedure. Averaged RAMSE estimates $\hat{\overline{L}}$ over a subset of the patches used for denoising $\mathcal{Q}''$ are computed for a set of candidate matrix aspect ratios $\gamma_{(j)}$ (which induce a set of patch radii). The candidate offering the minimum RAMSE $\hat{\gamma}$ is used for denoising.}
\label{fig:DPEP}
\end{figure}
\begin{itemize}
	\item \textbf{Patch extraction}. This step is analogous to that for actual denoising but a smaller number of patches $Q''$ is used to cover the FOV. Drawing a moderate number of patches will generally provide enough information for reliable global RAMSE estimates. Our in vivo tests have used a subsampling factor $T_{\gamma}=6$ in each direction. Using this factor,  the computation time of patch estimation has remained below the time for denoising.
	\item \textbf{SVD / ESD modelling}. These steps are the same as for actual denoising.
	\item \textbf{RAMSE estimation}. This step uses additional results provided in~\cite{Nadakuditi14} to obtain estimates of the asymptotic mean squared error (AMSE) of matrix recovery based on integral transforms of the ESD evaluated at the observed singular values above the detection threshold. Namely, for the optimal shrinker in~\eqref{ec:OSGE}
	\begin{equation}
	\label{ec:AMGE}
	\hat{L}^{f}_{\text{FROB}}(\hat{\mathbf{X}},\mathbf{X})=\sum_{r:\eta_{\mathbf{Y}r}>\eta_{\mathbf{W}}^{+}}\frac{1}{D_f(\eta_{\mathbf{Y}r})}-\hat{\eta}^2_{\mathbf{X}r}.
	\end{equation}
	To determine the optimal aspect ratio, the errors are normalized as follows
	\begin{equation}
	\hat{\overline{L}}=\frac{\displaystyle\sum_{q''\in\mathcal{Q}''}\hat{L}^{f}_{\text{FROB}}(\hat{\mathbf{X}}^{[q'']},\mathbf{X}^{[q'']})}{\displaystyle\sum_{q''\in\mathcal{Q}''}\|\hat{\mathbf{X}}^{[q'']}\|_2^2}.
	\end{equation}
	This feature is computed for a set of candidate $\gamma_{(l)}$ and the $\hat{\gamma}$ offering the minimum RAMSE $\hat{\overline{L}}$ is chosen for denoising. Importantly, these asymptotic error estimates can be used to assess the expected performance of different denoising alternatives in vivo without requiring a ground truth, which will be a key ingredient in~\S~\ref{sec:REVA}. Finally, note that the inverse of the RAMSE provides an estimate of the SNR after denoising.
\end{itemize}

\section{Validation and results}

\label{sec:REVA}

In this Section we validate the different contributions of our framework for the brain DWI domains and specific sequences described in~\S~\ref{sec:MATE} and using simulations. In~\S~\ref{sec:NOPR} we show the robustness of the noise measurement and propagation approach. Simulation-based experiments comparing our GSVS method with state of the art techniques are included in \S~\ref{sec:SIMU}. Improved recovery by using phase correction is tested in~\S~\ref{sec:PHST}. Effects and pertinence of applying our tools for generic noise models are studied in~\S~\ref{sec:GENO}. Finally, in~\S~\ref{sec:EDDM} we visually compare GSVS with the related MPPCA method in vivo.

\subsection{Explored applications}

\label{sec:MATE}

A series of experiments have been conducted including three distinct brain DWI application domains: high $b$-value exploration in adult brain imaging, multi-shell high angular resolution neonatal examinations, and multi-shell fetal imaging. All the volumes in the experiments have been acquired using a $3\,\mbox{T}$ \textsmaller{\textsc{Philips Achieva TX}} with $C=32$-channel coils. Written informed consent for each participant was obtained prior to them being scanned. Neonatal and fetal study procedures were reviewed and approved by the Riverside Research Ethics Committee (14/LO/1169). DWI data was acquired as part of a broader examination aimed to study brain development within the developing Human Connectome Project~\citep{dHCP}. Main sequence parameters are reported in Table~\ref{tab:SEPA} using a sequence design that allows for flexible sampling of the diffusion space, interleaved choice of the PE direction, free temporal arrangement of the sampling scheme, usage of SMS acceleration, and restart capacity~\citep{Hutter18}.

\begin{table}[!htb]
\begin{small}
\begin{center}
\begin{tabular}{c|c|c|c|c|c}
\hline
\multicolumn{6}{c}{Adult} \\ \hline\hline
Resolution ($\mbox{mm}^3$) & $\text{FOV}_{\text{rec}}$ ($\mbox{mm}^3$) & PE acc & SMS acc & PE shift & $b$-values ($\text{ms}/\mu\text{m}^2$) \\ \hline
$2.5\times 2.5\times 2.5$ & $240\times 240\times 100$ & $2$ & $2$ & $1/2$ & \vtop{\hbox{\strut $0$ / $0.1$ / $0.4$ / $0.9$ / $1.6$ / $2.5$}\hbox{\strut / $3.6$ / $4.9$ / $6.4$ / $8.1$ / $10$}} \\ \hline\hline
Slice distance ($\mbox{mm}$) & PEs & $T_{\text{R}}$ ($\mbox{s}$) & $T_{\text{E}}$ ($\mbox{ms}$) & PF & \# images/$b$-value \\ \hline
$2.5$ & AP & $6.9$ & $99$ & $1$ & \vtop{\hbox{\strut $15$ / $16$ / $19$ / $24$ / $31$ / $40$}\hbox{\strut / $51$ / $64$ / $79$ / $96$ / $115$}} \\ \hline\hline
\multicolumn{6}{c}{Neonatal} \\ \hline\hline
Resolution ($\mbox{mm}^3$) & $\text{FOV}_{\text{rec}}$ ($\mbox{mm}^3$) & PE acc & SMS acc & PE shift & $b$-values ($\text{ms}/\mu\text{m}^2$) \\ \hline
$1.5\times 1.52\times 3$ & $150\times 150\times 96$ & $1.19$ & $4$ & $1/2$ & $0.0$ / $0.4$ / $1.0$ / $2.6$ \\ \hline\hline
Slice distance ($\mbox{mm}$) & PEs & $T_{\text{R}}$ ($\mbox{s}$) & $T_{\text{E}}$ ($\mbox{ms}$) & PF & \# images/$b$-value \\ \hline
$1.5$ & LR/RL/AP/PA & $3.8$ & $90$ & $0.86$ & $20$ / $64$ / $88$ / $128$ \\ \hline\hline
\multicolumn{6}{c}{Fetal} \\ \hline\hline
Resolution ($\mbox{mm}^3$) & $\text{FOV}_{\text{rec}}$ ($\mbox{mm}^3$) & PE acc & SMS acc & PE shift & $b$-values ($\text{ms}/\mu\text{m}^2$) \\ \hline
$2.0\times 2.0\times 2.0$ & $200\times 312\times 108$ & $2$ & $2$ & $1/2$ & $0.0$ / $0.4$ / $1.0$ \\ \hline\hline
Slice distance ($\mbox{mm}$) & PEs & $T_{\text{R}}$ ($\mbox{s}$) & $T_{\text{E}}$ ($\mbox{ms}$) & PF & \# images/$b$-value \\ \hline
$2.0$ & AP & $6.2$ & $76$ / $137$ & $0.75$ & $16$ / $45$ / $80$ \\ \hline\hline
\end{tabular}
\end{center}
\end{small}
\caption{Sequence parameters for each of the DWI application domains. acc refers to acceleration factors.}
\label{tab:SEPA}
\end{table}

DWI recovery in these domains confronts different challenges. In the adult case, the high $b$-value DWI signal may fall below the Rician noise floor so the complex valued signal model seems advisable. In the neonatal case, we encounter a challenging acquisition protocol, where high spatial resolution DWI information has been acquired along four different PE directions interleaved in the diffusion space, using a SMS factor of $4$ and with large distortions due to low in-plane acceleration. In addition, these datasets are frequently hampered by motion artifacts, partly because the newborns have been scanned without sedation. Finally, in fetal examinations the targeted brain data is collected from locations that may be far away from the coils, with consequent low SNR levels. In addition, there is breathing motion and perhaps fetal motion throughout the acquisition, so that retrieval and representation of DWI information becomes extremely challenging. Two $T_{\text{E}}$ values are reported for this dataset as it has been acquired using a dual hybrid spin echo and field echo sequence where the dead time after the application of the spin echo is used to collect some field echo information that is later used to correct for dynamic main magnetic field variations, which will be the subject of a future publication. Both echoes are acquired with the exact same encoding structure, so we can easily incorporate the multiple echo information to our denoising technique, as previously proposed in~\cite{Bydder06}.

\subsection{Propagation of noise measures}

\label{sec:NOPR}


When the noise level is not available before denoising, different techniques have been proposed for its estimation. In~\cite{Veraart16b} the underlying matrix rank $\hat{R}$ and the noise level $\hat{\sigma}$ are simultaneously estimated using statistics on the expectation of the Mar\v{c}enko-Pastur distribution for noise matrices with additive i.i.d. entries. This estimator is identified hereafter as EXP1. The procedure is summarized by~\eqref{ec:NET1} within Table~\ref{tab:NETC}; increasing signal ranks $R$ are generated until a given criterion for the rank estimate depending on standardized and decreasingly sorted eigenvalues is satisfied, with the standardization performed using the estimate for $\sigma^2$ at the current signal rank, $\tilde{\sigma}^2(R)$. We have studied this estimator and observed a biased behaviour when applied to close to square matrices, so we propose a simple modification that we identify as EXP2 in~\eqref{ec:NET1}. On the other hand, in~\cite{Gavish14} a simple estimator is proposed where the standard deviation of noise is computed as the ratio of the median of the observed eigenvalues, $\eta^2_{\mathbf{Y}M/2}$, and the median of the Mar\v{c}enko-Pastur law, $\theta^{2}_{M/2}$, as shown in~\eqref{ec:NET2}. This estimator will be identified as MED.
\begin{table}[!htb]
\begin{center}
\begin{tabular}{c|c}
Technique / Reference & Noise estimate \\ \hline
\begin{minipage}{.25\textwidth}\begin{center}\small{Spectral expectation\\\cite{Veraart16b}}\end{center}\end{minipage} & 
\begin{minipage}{0.6\textwidth}\small{\begin{equation}\label{ec:NET1}
\begin{split}\tilde{\sigma}^2_{\text{EXP1}}(R)=&\frac{\sqrt{N}(\eta^2_{\mathbf{Y}R+1}-\eta^2_{\mathbf{Y}M})}{4\sqrt{M-R}}\\
\tilde{\sigma}^2_{\text{EXP2}}(R)=&\frac{\sqrt{N-R}(\eta^2_{\mathbf{Y}R+1}-\eta^2_{\mathbf{Y}M})}{4\sqrt{M-R}},\\
\hat{R}_{\text{EXP}\{1,2\}}=&\inf\left\{R\displaystyle\mathrel{\Bigg|}\sum_{r=R+1}^{M}\eta^2_{\mathbf{Y}r}\geq (M-R)\tilde{\sigma}^2_{\text{EXP\{1,2\}}}(R)\right\}\\
\hat{\sigma}^2_{\text{EXP\{1,2\}}}=&\tilde{\sigma}^2(\hat{R}_{\text{EXP\{1,2\}}})
\end{split}
\end{equation}}\end{minipage} \\ \hline
\begin{minipage}{.25\textwidth}\begin{center}\small{Spectral median\\\cite{Gavish14}}\end{center}\end{minipage} & 
\begin{minipage}{0.55\textwidth}\small{\begin{equation}\label{ec:NET2}
\hat{\sigma}^2_{\text{MED}}=\displaystyle\frac{\eta^2_{\mathbf{Y}M/2}}{\theta^{2}_{M/2}}
\end{equation}}\end{minipage} \\ \hline
\end{tabular}
\end{center}
\caption{Considered noise estimation techniques.}
\label{tab:NETC}
\end{table}

To show the correctness of noise propagation through the reconstruction as well as the limitations of these data-based noise estimation methods, we use the full Fourier adult dataset (PF factor of $1$ in Table~\ref{tab:SEPA}). In this case, noise is captured by~\eqref{ec:YPIN}. Strictly speaking, the noise level within a given patch is not unique as it changes according to the spatial variation of the receiver field and the g-factor profiles, with the latter potentially inducing abrupt changes in the noise properties of neighboring locations. To avoid this inherent limitation of data-based noise level estimation when operating in signal units, our pipelines are modified in this experiment so that the data is standardized by the application of a whitening transformation $\boldsymbol{\Upsilon}_M$ that normalizes by the noise factors. After applying this transformation, optimal asymptotic loss guarantees are pursued on the invariant Frobenius loss~\citep{Canu17}, $L_{\text{FROB},\boldsymbol{\Upsilon}_M}(\hat{\mathbf{X}},\mathbf{X})=\|\boldsymbol{\Upsilon}_M(\hat{\mathbf{X}}-\mathbf{X})\|_2^2$. If our noise measurement and propagation model is correct, the estimated noise level after standardization should be $\hat{\sigma}=1$ everywhere. We check whether this is the case by running the estimators in Table~\ref{tab:NETC} for reconstructions of noise-only data with the noise injected in place of the original data after the channel noise standardization described in~\S~\ref{sec:MNOI}. These results are compared with the estimations obtained when reconstructing the original signal plus noise data.

\begin{figure}[!htb]
\begin{minipage}{0.59\textwidth}
\begin{center}
\includegraphics[width=\textwidth]{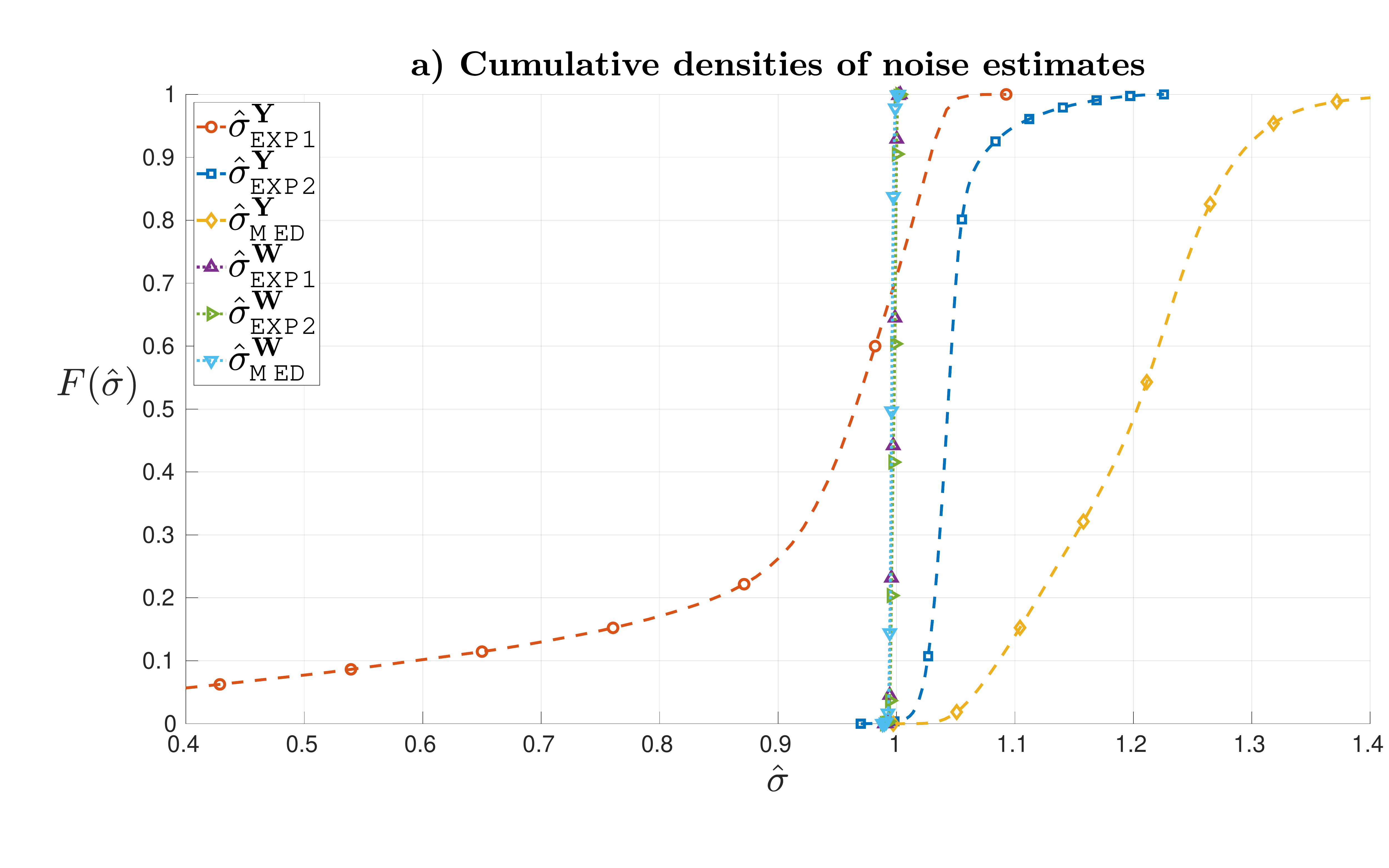}
\end{center}
\end{minipage}
\begin{minipage}{0.39\textwidth}
\begin{center}
\includegraphics[width=0.49\textwidth]{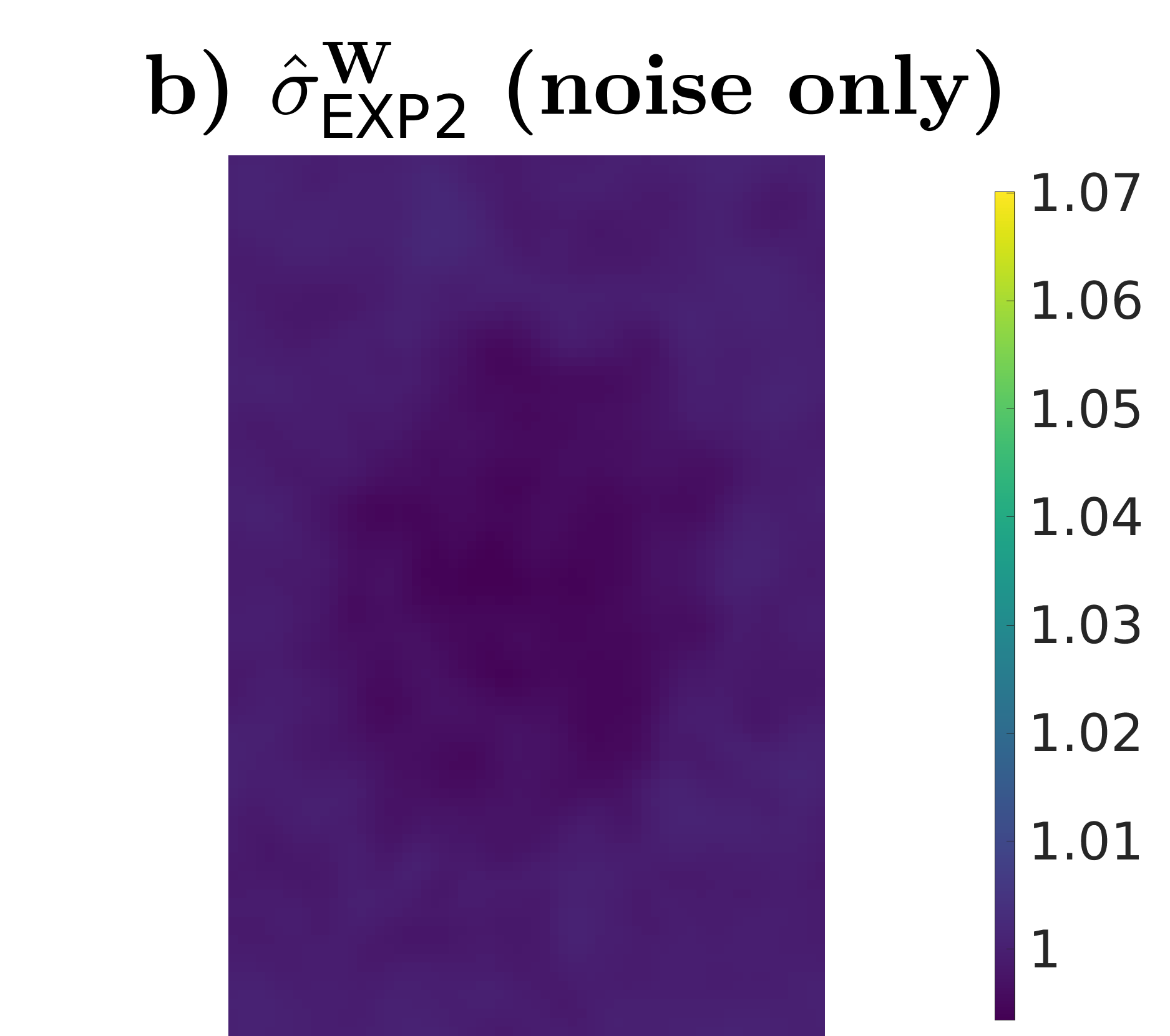}
\includegraphics[width=0.49\textwidth]{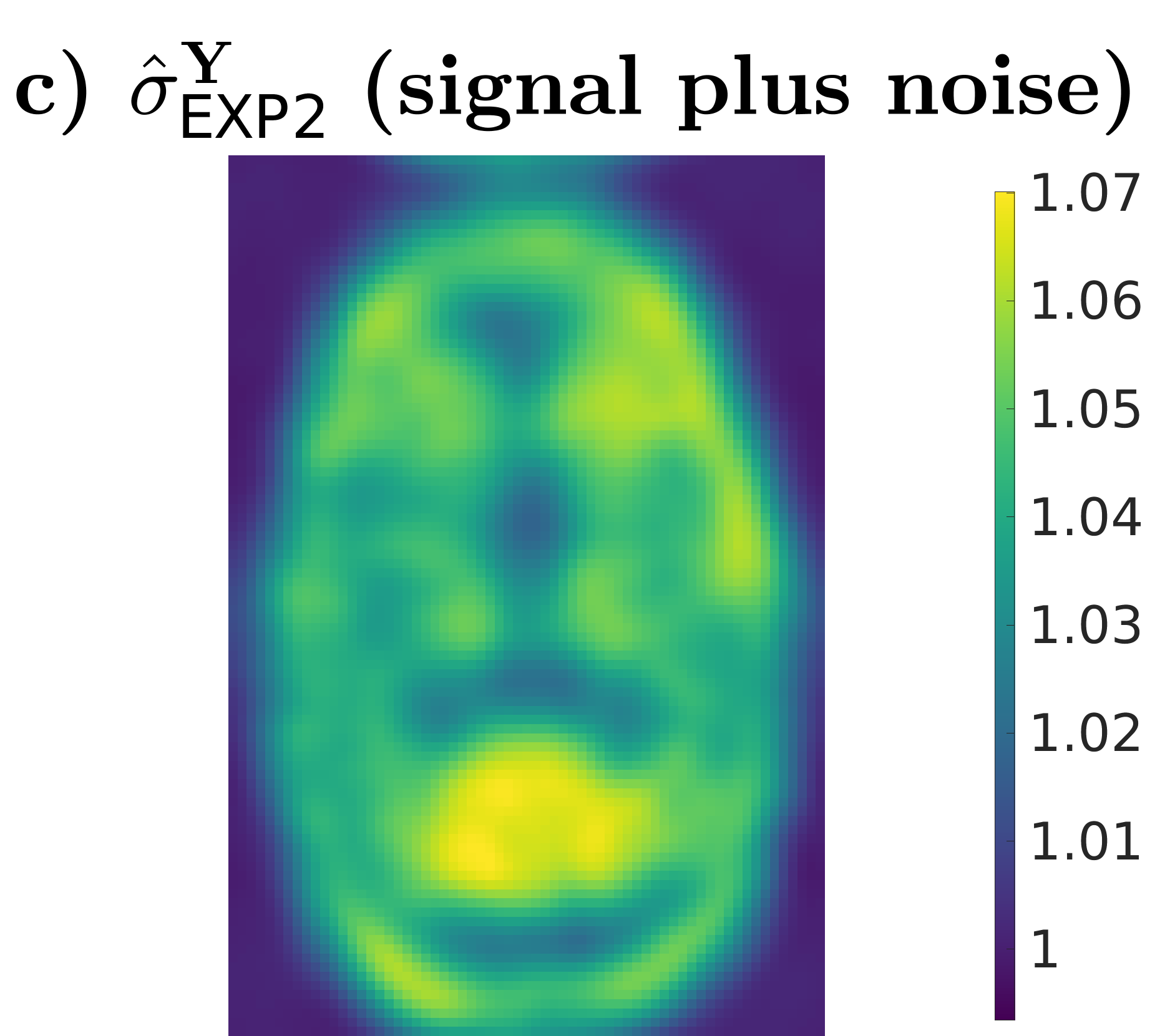}\\
\includegraphics[width=0.49\textwidth]{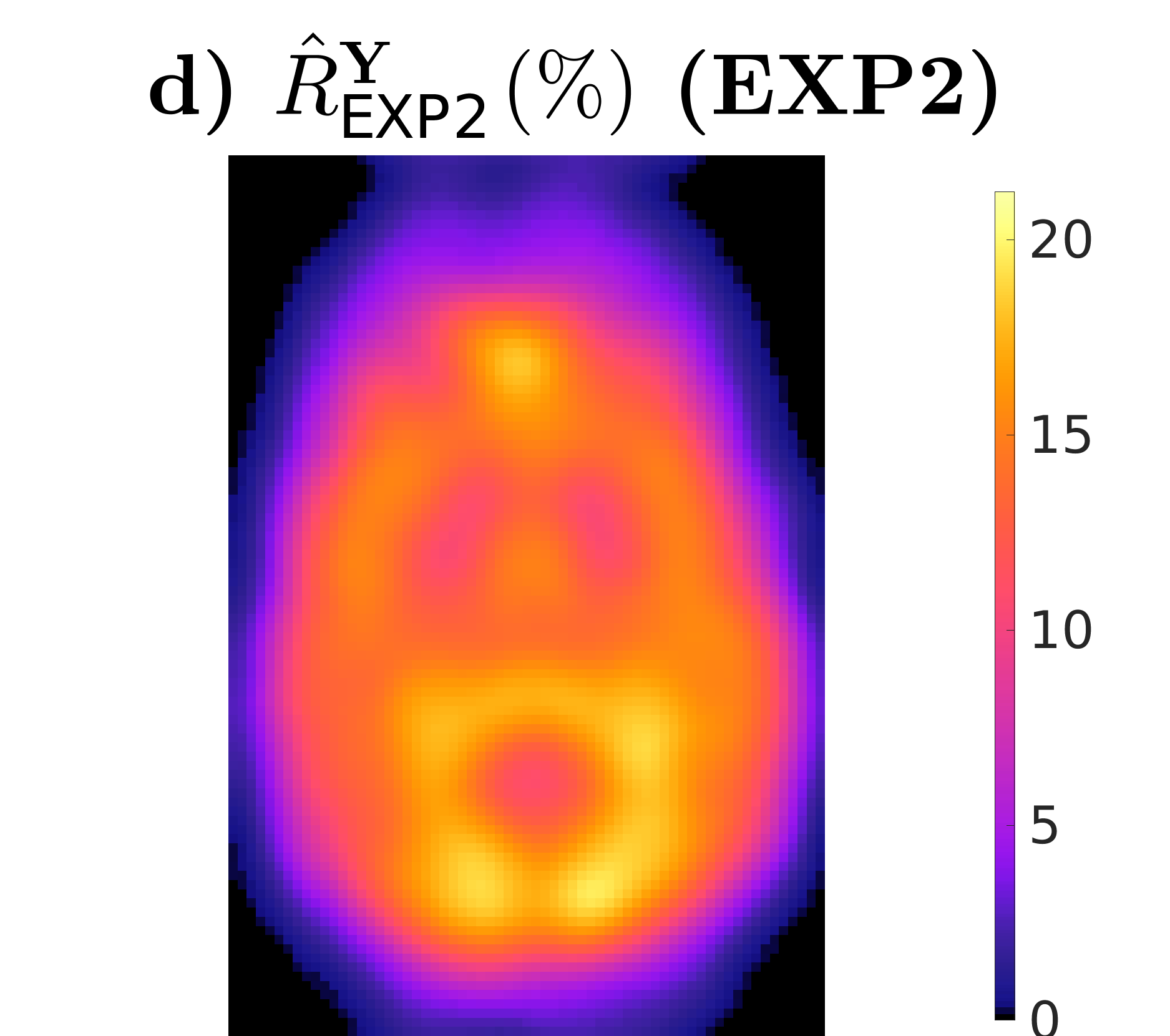}
\includegraphics[width=0.49\textwidth]{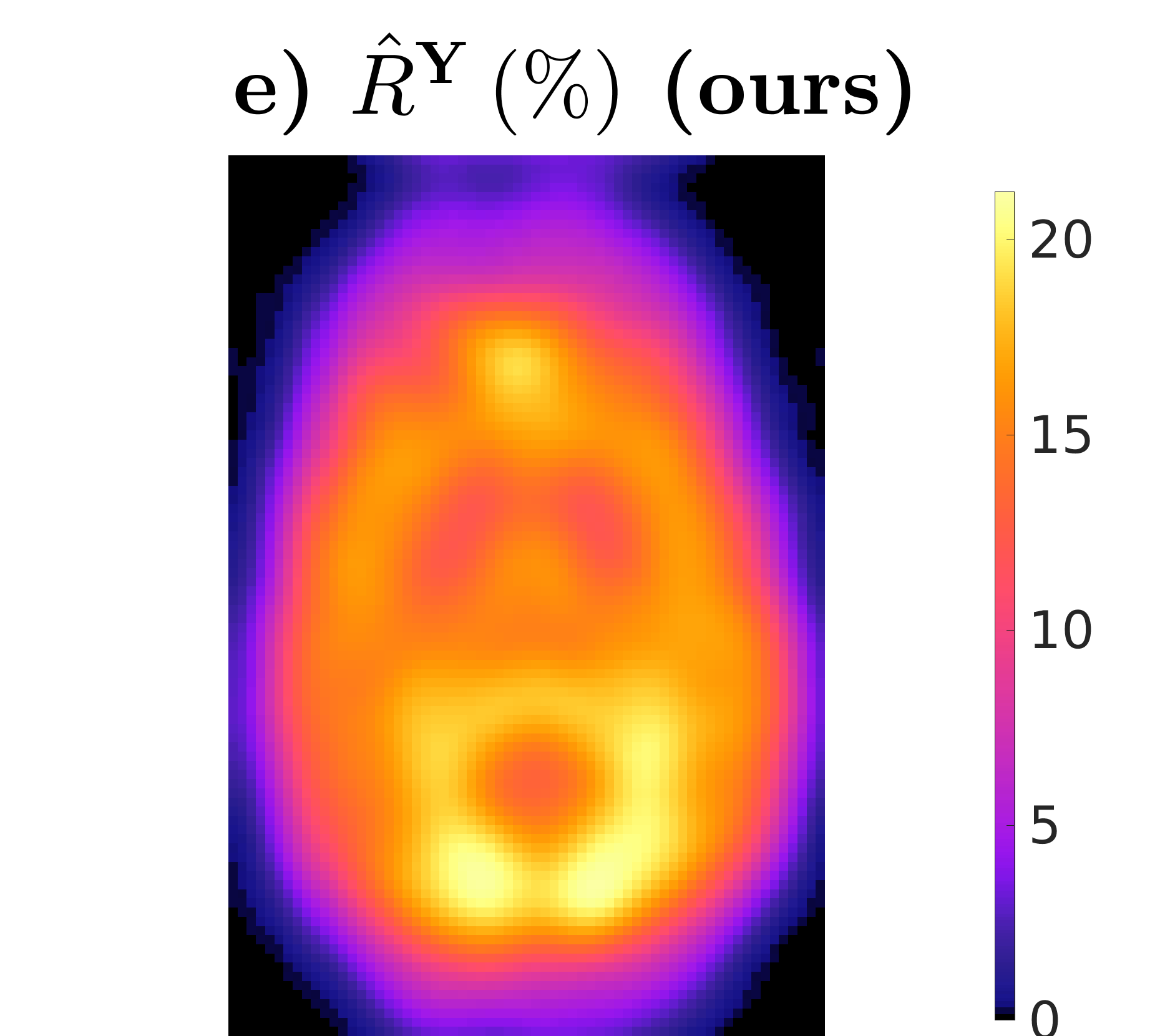}
\end{center}
\end{minipage}
\caption{\textbf{a)} Cumulative spatial distributions of noise estimates for different estimators without ($\hat{\sigma}^{\mathbf{W}}$) and with ($\hat{\sigma}^{\mathbf{Y}}$) signal perturbation. \textbf{b,c)} Spatial maps of noise for estimator EXP2 \textbf{b)} without and \textbf{c)} with signal perturbation. \textbf{d,e)} Rank estimates $\hat{R}$, given as a percentage of the number of volumes $N$, for estimators \textbf{d)} EXP2 and \textbf{e)} ours. All results correspond to the RAMSE-optimal aspect ratio $\hat{\gamma}=0.85$.}
\label{fig:NOPS}
\end{figure}

Fig.~\ref{fig:NOPS}a compares the cumulative histograms of noise estimates at the different spatial locations. Results show that when estimating noise levels in the absence of additive signal perturbations (cases $\hat{\sigma}^{\mathbf{W}}$) the distribution of estimates takes the value $1$ everywhere, with only minor numerical deviations. This shows both the consistency of the estimators in Table~\ref{tab:NETC} and the correctness of our noise propagation method. However, in the presence of signal perturbations (cases $\hat{\sigma}^{\mathbf{Y}}$), the estimators in Table~\ref{tab:NETC} start to deviate from their asymptotic behaviour, with positive biases for EXP2 and MED and potentially larger dispersions for EXP1, which shows the limitations of data-based noise estimation techniques. Figs.~\ref{fig:NOPS}b,c show exemplary spatial maps with the estimated noise levels for the noise-only data and the signal plus noise data using the shape corrected version of~\eqref{ec:NET1}, EXP2, which has been judged to provide the most benign dispersion and bias tradeoff according to Fig.~\ref{fig:NOPS}a. On the other hand, Figs.~\ref{fig:NOPS}d,e compare the estimated ranks when using EXP2 and our noise propagation and asymptotic random matrix rank estimation. The results are in agreement with the elementary interplay where noise overestimation of EXP2 (Fig.~\ref{fig:NOPS}c) is accompanied by lower rank estimates (Fig.~\ref{fig:NOPS}d).

\subsection{Simulation-based validation}

\label{sec:SIMU}

A set of synthetic experiments have been conducted to quantitatively assess the performance of our method. Namely, we have used the \texttt{\emph{phantom$\alpha$s}} software~\citep{Caruyer19} to synthesize the $550$ diffusion gradients corresponding to the adult protocol in Table~\ref{tab:SEPA} using the default diffusion parameters. Realistic SNR levels have been simulated by median-based harmonization of the $b=0$ intensity levels in the phantom to those of the adult scan and noise generation according to the noise covariance after reconstruction. As we have already mentioned, most of current advanced denoising techniques are proposed for magnitude-only data, which complicates comparisons with the complex-domain methodology proposed here. This difficulty has been circumvented by performing comparisons within an idealized scenario where we assume that real-only information with complete restoration of Gaussianity has been made available by a given bias correction technique. Two state of the art methods have been chosen for comparison, the related MPPCA in~\cite{Veraart16b} and the non local spatial and angular matching (NLSAM) non-local means based approach in~\cite{St-Jean16}, to our knowledge the most recent non-local means approach for DWI with an available on-line implementation. NLSAM allows to input a noise standard deviation map, which enables more direct comparisons to our methodology. This is not the case for MPPCA, so we have resorted to the EXP2 noise estimation in~\eqref{ec:NET1}. To discern the impact of noise estimation versus noise modeling, we have also applied our technique using noise maps from the EXP2 patch estimates instead of the noise characterization derived from the reconstruction, which we term as the GSVS$\hat{\sigma}$ alternative.

\begin{figure}[!htb]
\begin{minipage}{0.49\textwidth}
\begin{center}
\includegraphics[width=\textwidth]{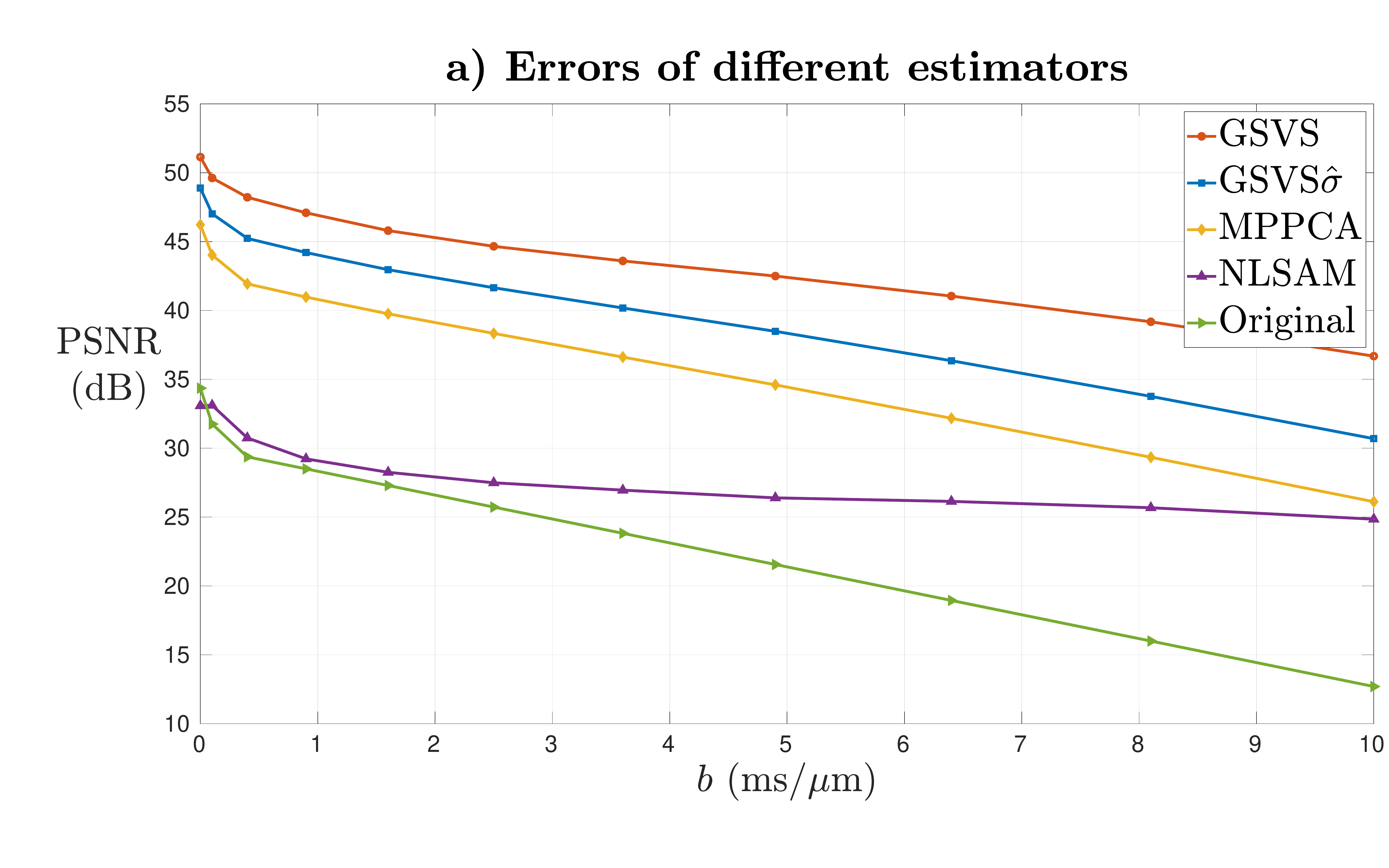}
\end{center}
\end{minipage}
\begin{minipage}{0.49\textwidth}
\begin{center}
\includegraphics[width=\textwidth]{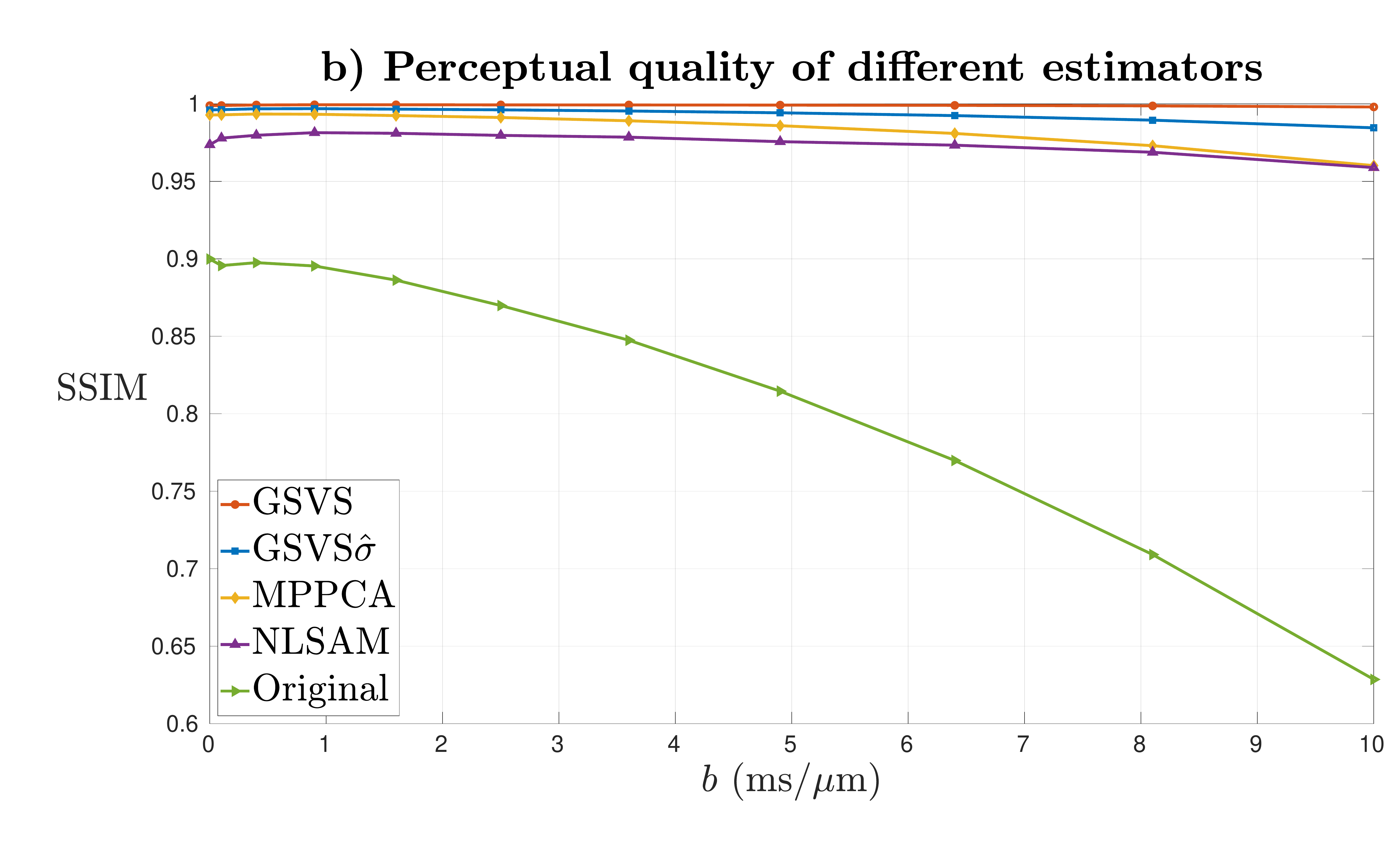}
\end{center}
\end{minipage}\\
\begin{minipage}{0.49\textwidth}
\begin{center}
\includegraphics[width=0.65\textwidth]{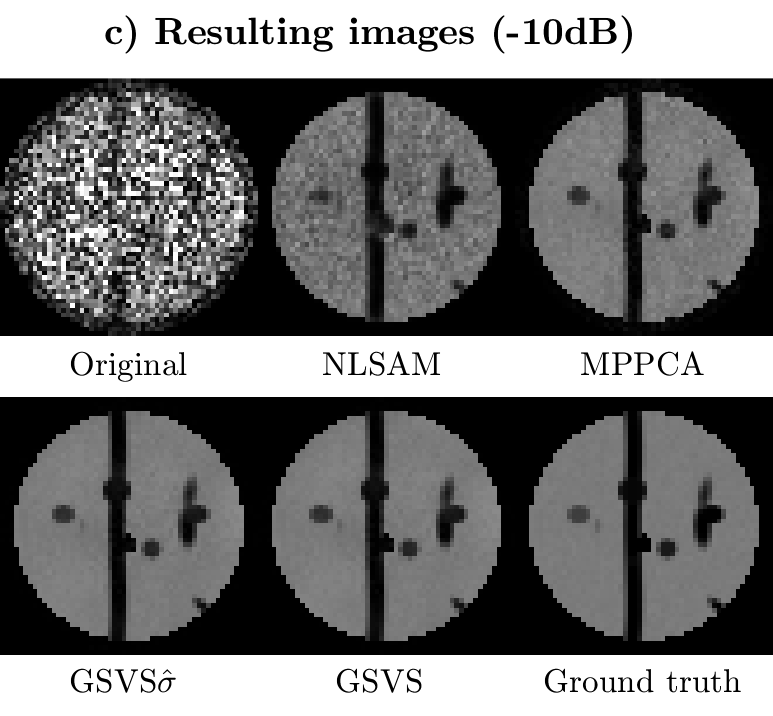}
\end{center}
\end{minipage}
\begin{minipage}{0.49\textwidth}
\begin{center}
\includegraphics[width=\textwidth]{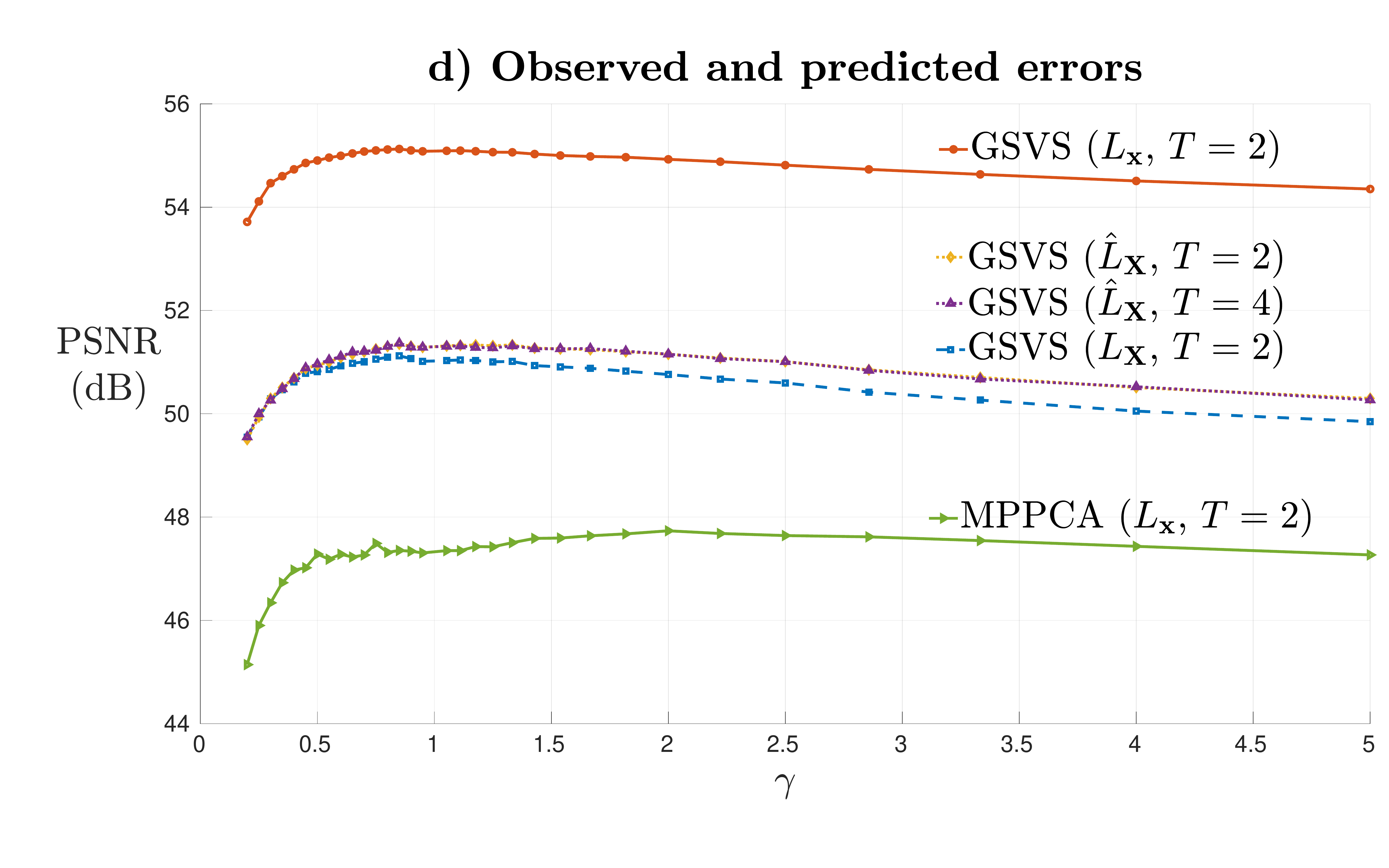}
\end{center}
\end{minipage}
\caption{Comparison of different denoising approaches on a synthetic phantom. \textbf{a)} PSNR and \textbf{b)} SSIM for the compared approaches and the original data at a baseline SNR that matches a real acquisition. \textbf{c)} Exemplary images with $10\,\mbox{dB}$ SNR attenuation with respect to baseline when synthesizing the data. \textbf{d)} Measured PSNR in the resulting image (solid), together with measured (dashed) and predicted (dotted) PSNR in the matrices for different methods and patch subsampling factors.}
\label{fig:SIMU}
\end{figure}

Figs.~\ref{fig:SIMU}a,b display the peak SNR (PSNR) and structural similarity index (SSIM) perceptual metric~\citep{Wang04} for each method as functions of the $b$-values in the simulated sequence. Results show that the GSVS approach ranks the best for all $b$-values both in terms of PSNR and SSIM. On average, the PSNR of the GSVS method improves by $7.82\,\mbox{dB}$ when compared to MPPCA, with $3.96\,\mbox{dB}$ basically attributable to noise modeling (difference with GSVS$\hat{\sigma}$) and the remaining $3.86\,\mbox{dB}$ due to the refinements in the statistical characterization of the problem. The difference rises to $16.32\,\mbox{dB}$ and $20.81\,\mbox{dB}$ when comparing to NLSAM and synthesized data respectively. In Fig.~\ref{fig:SIMU}c we show a visual example for a slice with $b=10\,\mbox{ms/}\mu\mbox{m}$ of an analogous experiment where the SNR of the original data is attenuated by $10\,\mbox{dB}$, as this SNR regime allows for easier perception of the differences between methods. Despite the original simulated data being severely degraded by noise, the underlying structures are visible after applying any of the denoising strategies but stronger noise suppression and fewer artifacts are noticeable when using the GSVS or GSVS$\sigma$ approaches. Total computing time of our method has been of $194\,\mbox{s}$, with $34\,\mbox{s}$ for patch size estimation and $160\,\mbox{s}$ for actual denoising, which compares favourably with the $6836\,\mbox{s}$ ($8$-core multiprocessing) employed by NLSAM and is similar to the MPPCA time. In Fig.~\ref{fig:SIMU}d we look at the assumptions of the patch size estimation procedure as well as those of the methodology for in vivo comparisons. Using a subsampling factor of $T=2$ as a reference, we compare the global PSNR $L_{\mathbf{x}}$ of GSVS (red solid) and MPPCA (green solid), the global PSNR of GSVS without patch overlap $L_{\mathbf{X}}$ (blue dashed) and the predicted global PSNR of GSVS $\hat{L}_{\mathbf{X}}$ both for $T=2$ (dotted yellow) and $T=4$ (dotted magenta) obtained by the AMSE in~\eqref{ec:AMGE}. We observe that the profiles of the curves are similar for the range of compared aspect ratios $\gamma$ with the GSVS outperforming MPPCA across the whole range. PSNR improvement due to patch assembling is also noticeable for all compared $\gamma$. The PSNR predictions generally provide a good approximation of the observed errors before patch assembling with estimates being practically equivalent for $T=2$ and $T=4$ as these are aggregates averaged across a large number of patches. However, there is a certain overestimation of errors for large aspect ratios. Closer inspection has revealed that this is due to relatively larger impact of deviations from the asymptotic assumptions for large $\gamma$. This limitation, probably related to the number of variables $M$ being bigger than the number of observations $N$ has motivated constraining the patch size estimation to $\gamma<1$ for in vivo applications. Finally, the errors are relatively stable around the optimal $\gamma$ so approximate techniques for patch size estimation seem to generally suffice in practice.

\subsection{Effect of phase correction}

\label{sec:PHST}

\begin{figure}[!htb]
\begin{minipage}{0.55\textwidth}
\begin{center}
\includegraphics[width=\textwidth]{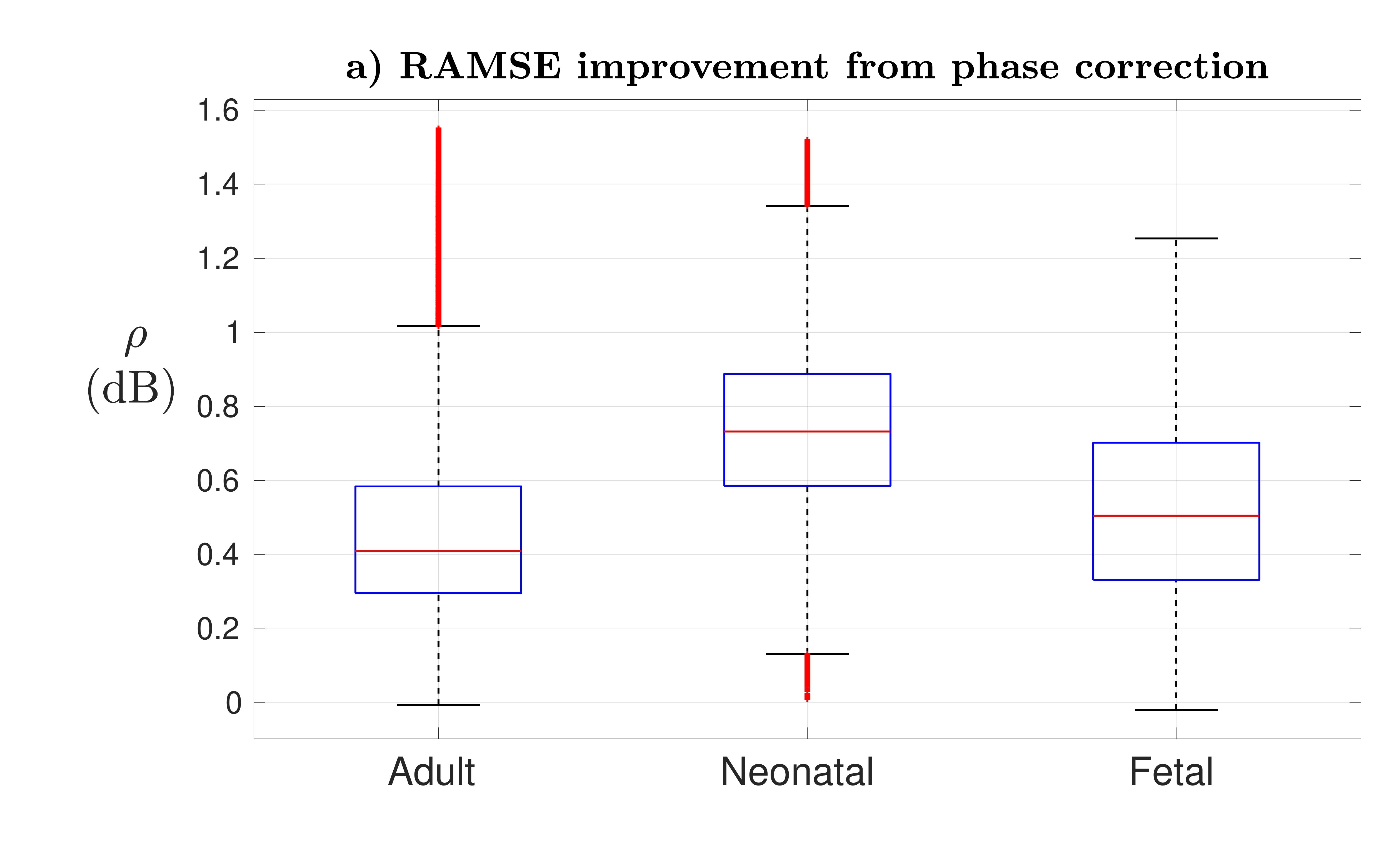}
\end{center}
\end{minipage}
\begin{minipage}{0.44\textwidth}
\begin{center}
\includegraphics[width=0.32\textwidth]{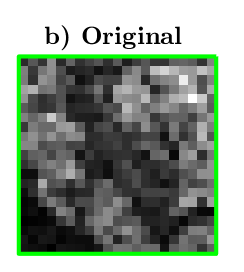}
\includegraphics[width=0.32\textwidth]{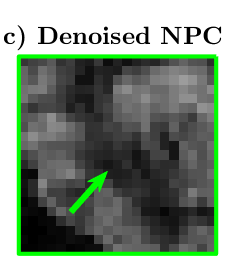}
\includegraphics[width=0.32\textwidth]{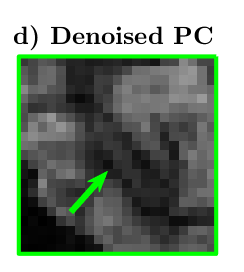}\\
\includegraphics[width=0.32\textwidth]{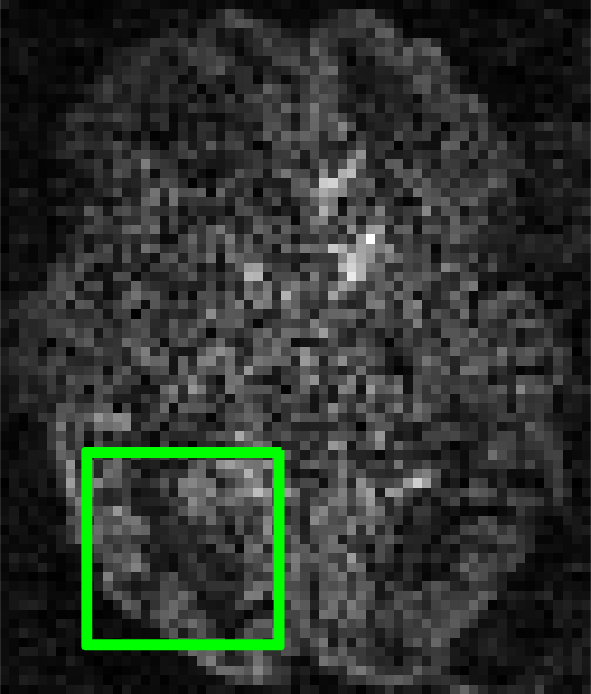}
\includegraphics[width=0.32\textwidth]{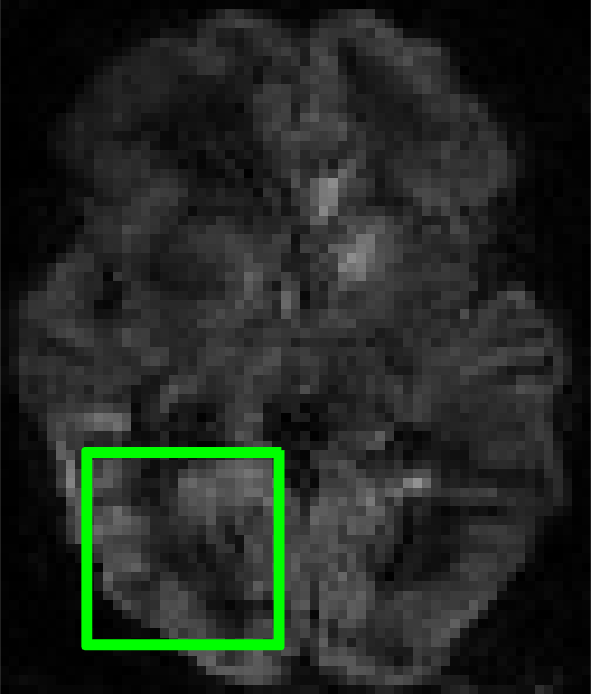}
\includegraphics[width=0.32\textwidth]{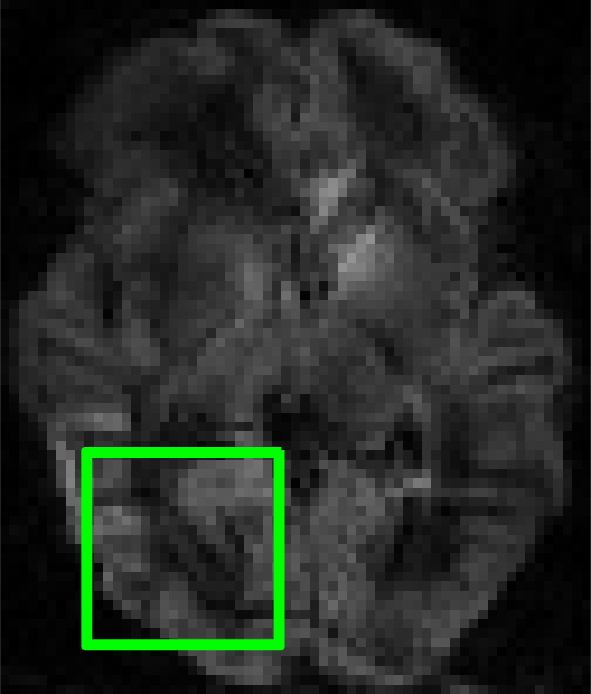}
\end{center}
\end{minipage}
\caption{\textbf{a)} RAMSE improvement ratio $\rho$ of denoising with phase correction (PC) versus denoising without phase correction (NPC). \textbf{b-d)} NPC and PC denoising results for an exemplary slice at $b=2.6\text{ms}/\mu\text{m}^2$ in the neonatal case: \textbf{b)} original data, \textbf{c)} NPC denoising, \textbf{d)} PC denoising . Same estimated aspect ratios $\hat{\gamma}$ have been obtained for NPC and PC denoising, respectively of $0.85$, $0.8$ and $0.9$ for adult, neonatal and fetal data. Mean inverse RAMSEs with phase correction have been $20.52$, $20.17$ and $12.32\,\mbox{dB}$ respectively for the adult, neonatal and fetal cases.}
\label{fig:DPCQ}
\end{figure}

In Fig.~\ref{fig:DPCQ}a we show the spatial distributions of the improvement ratio $\rho$ in the RAMSE when using the full method (PC) versus removing the phase correction in~\S~\ref{sec:TALD} (NPC), which is computed as\newline $\rho=-10\log_{10}\hat{\overline{L}}_{\text{PC}}/\hat{\overline{L}}_{\text{NPC}}$. Plots are generated using brain-only information and they include the adult case and exemplary neonatal and fetal studies drawn at random. A reduction of the RAMSE is observed in all tested cases when using phase correction, with improvements ranging from around $0$ to $1.5\,\mbox{dB}$ depending on the location. In Figs.~\ref{fig:DPCQ}b-d we illustrate with comparisons of original data, results of denoising without phase correction and results of denoising with phase correction in an exemplary subject, slice and orientation within the highest $b$-value shell for the neonatal case. Despite the estimator operates on complex data, for the sake of economy we only provide magnitude displays of the results. Emergence of plausible anatomical features with better preserved resolution can be observed when using phase correction, for instance in the area of the posterior thalamic radiation indicated by the arrow.

\subsection{Denoising with a generic noise model}

\label{sec:GENO}

\begin{figure}[!htb]
\begin{minipage}{0.31\textwidth}
\includegraphics[width=\textwidth]{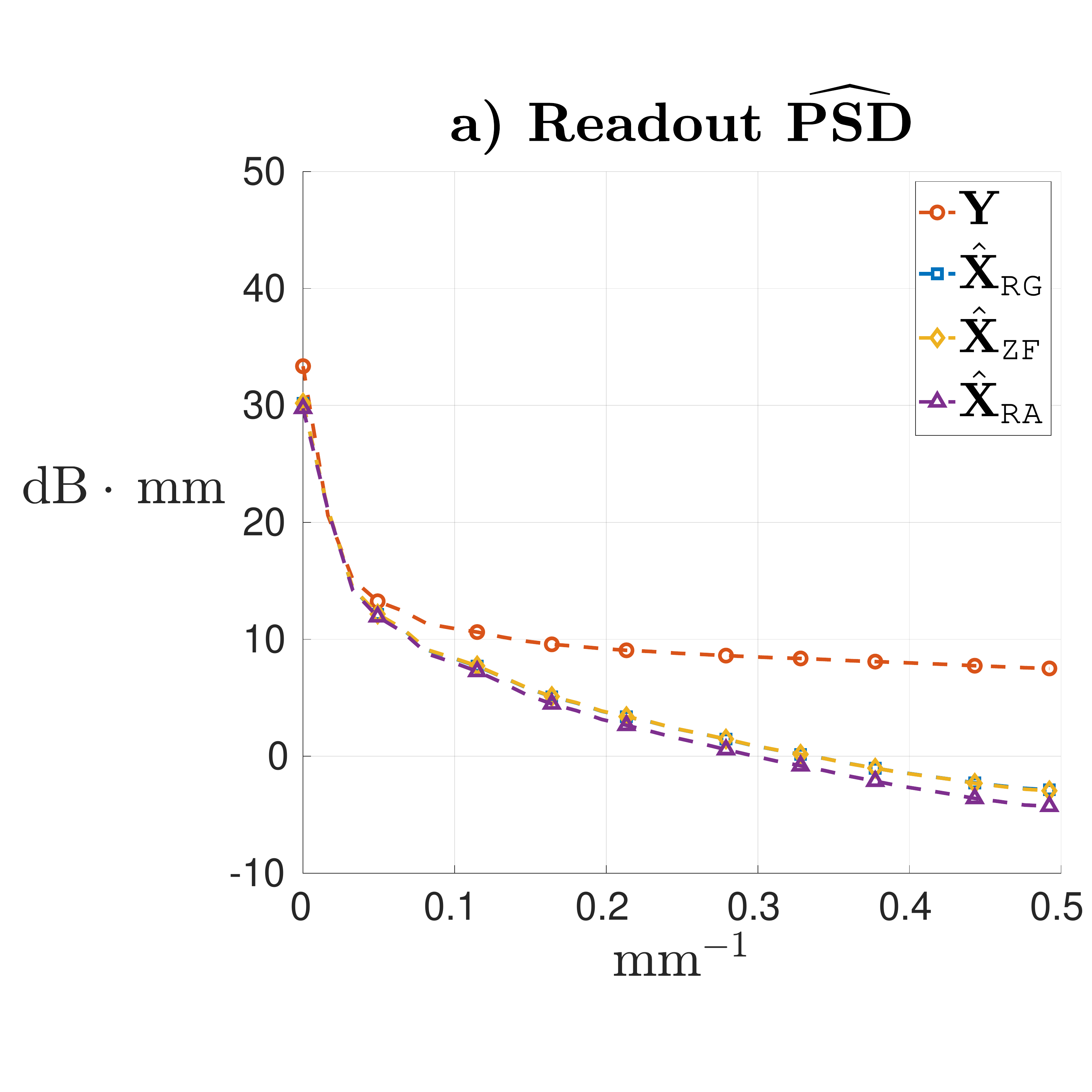}
\end{minipage}
\begin{minipage}{0.31\textwidth}
\includegraphics[width=\textwidth]{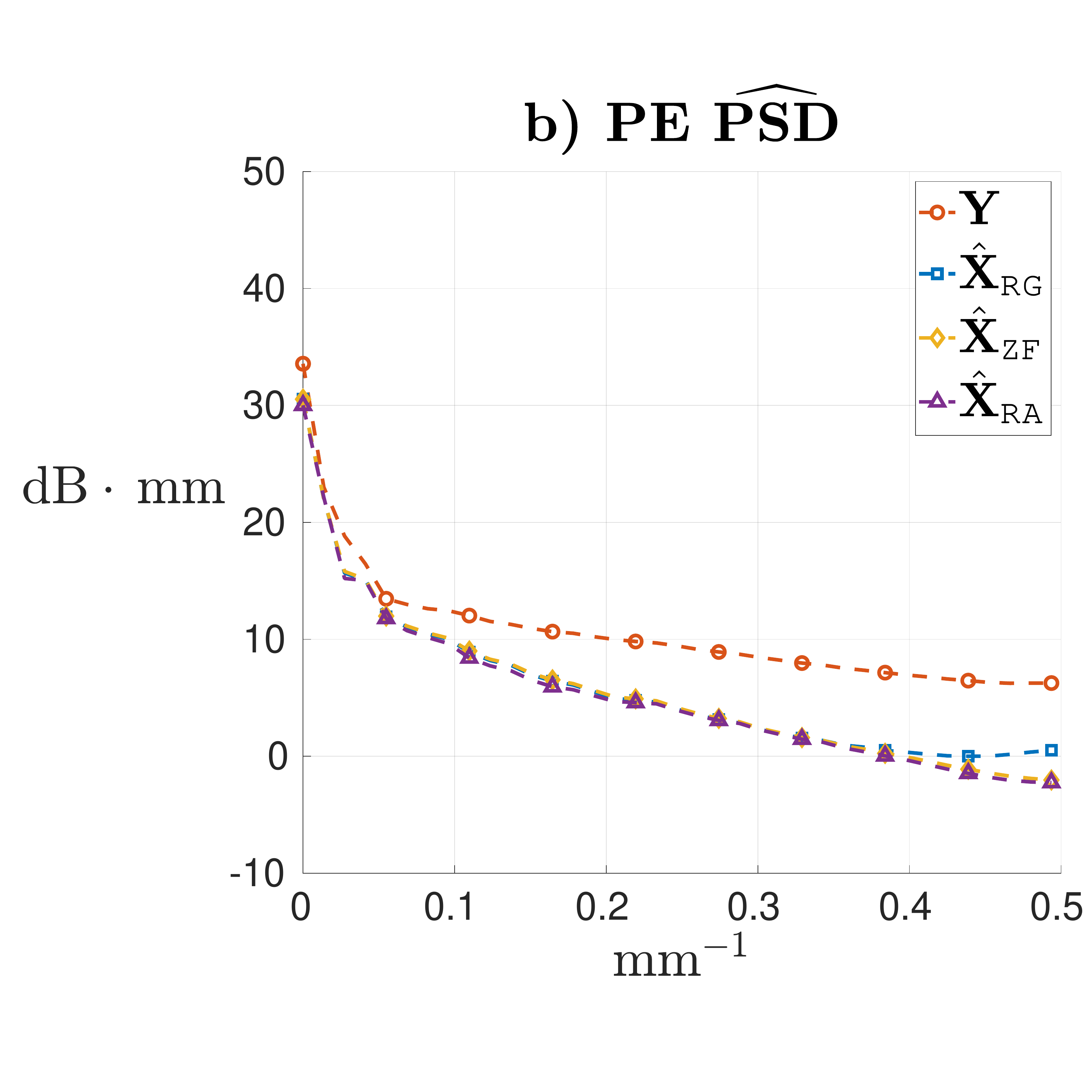}
\end{minipage}
\begin{minipage}{0.37\textwidth}
\includegraphics[width=0.49\textwidth]{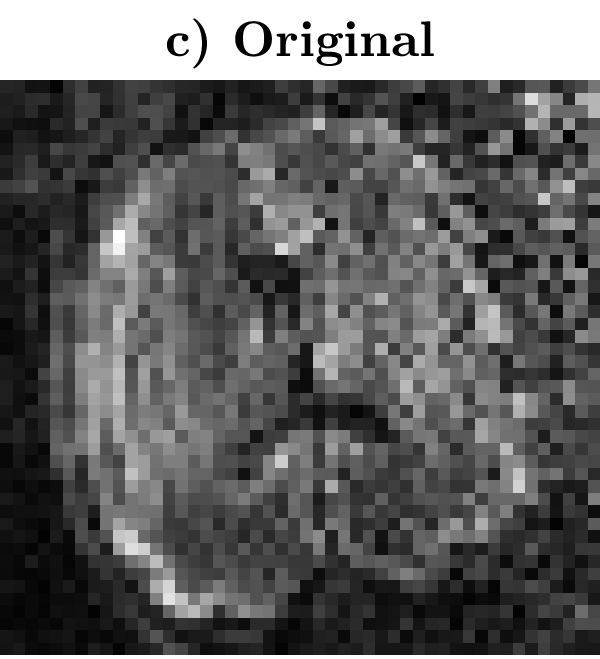}
\includegraphics[width=0.49\textwidth]{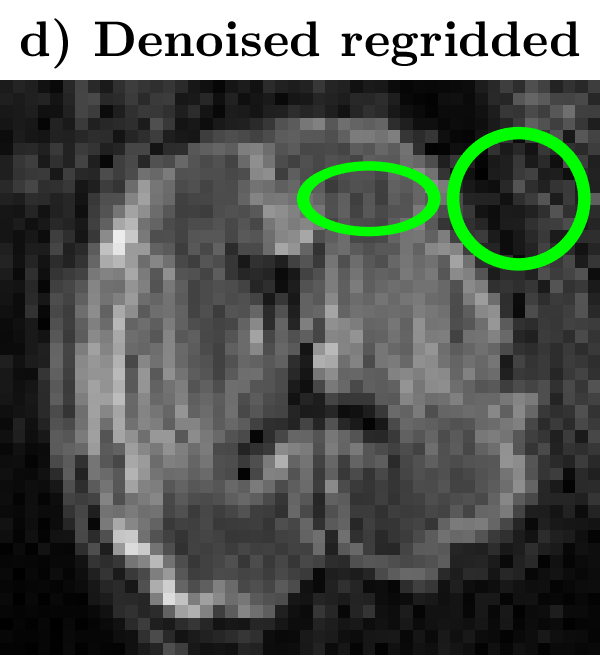}\\
\includegraphics[width=0.49\textwidth]{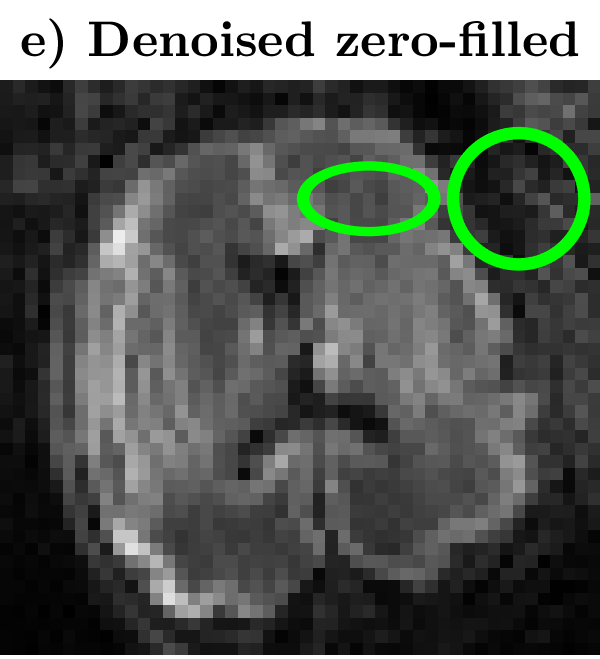}
\includegraphics[width=0.49\textwidth]{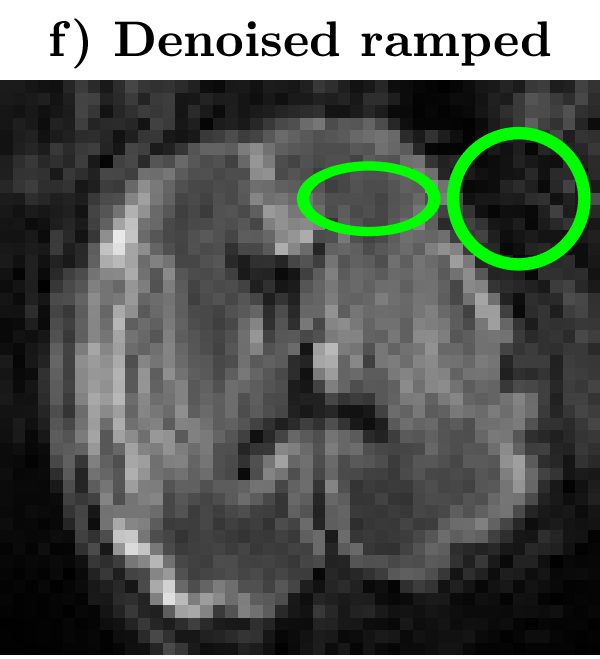}
\end{minipage}
\caption{\textbf{a,b)} PSD for non-denoised data, $\widehat{\text{PSD}}^{\mathbf{Y}}$, and the different denoising alternatives, $\widehat{\text{PSD}}^{\mathbf{X}_{\text{RG}}}$ (regridded), $\widehat{\text{PSD}}^{\mathbf{X}_{\text{ZF}}}$ (zero-filled) and $\widehat{\text{PSD}}^{\mathbf{X}_{\text{RA}}}$ (ramped), along the \textbf{a)} readout and \textbf{b)} PE directions. \textbf{c-f)} Resulting images \textbf{c)} before denoising, and when denoising using the \textbf{d)} regridded ($\hat{\gamma}=0.65$), \textbf{e)} zero-filled ($\hat{\gamma}=0.85$) and \textbf{f)} ramped ($\hat{\gamma}=0.9$) approaches.}
\label{fig:DGNM}
\end{figure}

The flexibility of the denoising scheme is illustrated by its application under different reconstruction operators and associated noise models. In a first experiment we use the fetal data to compare three approaches for PF reconstruction. In the first approach, hereafter referred to as regridded, the reconstructed PF data is regridded so that only the sampled area of the spectrum is back-transformed and consequently the data is represented in a coarser grid in the PE direction but with preserved local noise independence. In this case, the noise model follows~\eqref{ec:YPIN} and the spatial patches are drawn elliptically rather than spherically. After denoising, the data is zero-filled and ramp-filtered as in~\cite{Noll91} to retrieve the desired grid and resolution. As for the second approach, referred to as zero-filled, $\mathbf{G}$ in~\eqref{ec:YHIN} takes the role of a zero filling filter, $\mathbf{G}_{\text{ZF}}$, with a ramp filter being applied after denoising. The third approach, referred to as ramped, applies the zero filling and ramp filters before denoising, so $\mathbf{G}=\mathbf{G}_{\text{RA}}$. Figs.~\ref{fig:DGNM}a,b show, for the spin echo data at the highest $b$-value shell, the power spectral density (PSD) of the recovered magnitude signal in both the readout and PE directions for the three denoising approaches and the original data. The PSD along a particular dimension is estimated by averaging along the remaining dimensions and the diffusion orientations. The PSDs generally follow a plausible linear decay after denoising. Overall, the ramped approach provides stronger noise suppression than the other two in the readout direction (as given by more pronounced PSD decay ratios) and similar suppression than the zero-filled approach in the PE direction, with weaker suppression and implausible shape for the regridded PE PSD at high frequencies. These results are in agreement with the implicit SNR-resolution tradeoffs of the different models used for denoising, with lower SNR of ramped when compared to zero-filled~\citep{Noll91} and larger targeted resolution of zero-filled versus regridded. The low rank prediction adapts to the different SNR-resolution tradeoffs as reflected by the differences in resulting PSD profiles. The denoised images are compared in Figs.~\ref{fig:DGNM}c-f, showing the original data (Fig.~\ref{fig:DGNM}c) and the data denoised with the considered PF approaches. Regridded denoising (Fig.~\ref{fig:DGNM}d) produces some spurious oscillations along the PE (horizontal) direction (see the area enclosed by the ellipse), which agrees with the shape of the PE PSD response at high frequencies. As for the zero-filled (Fig.~\ref{fig:DGNM}e) and ramped (Fig.~\ref{fig:DGNM}f) results, differences are more subtle but stronger background noise suppression can be appraised in the ramped case (see the area enclosed by the circle), which seems also in agreement with the slightly stronger decay of the readout PSD for this approach. These experiments show that our denoising method provides flexible ways of balancing the SNR-resolution tradeoff in PF reconstructions considering, for instance, the noise amplification introduced when ramp filtering in homodyne reconstructions. They also show that while it is possible to reduce the problem to one with independent noise samples along the local patches by manipulating the reconstruction grid, this has an impact on the results at the prescribed grid and resolution.

\begin{figure}[!htb]
\begin{minipage}{0.49\textwidth}
\includegraphics[width=\textwidth]{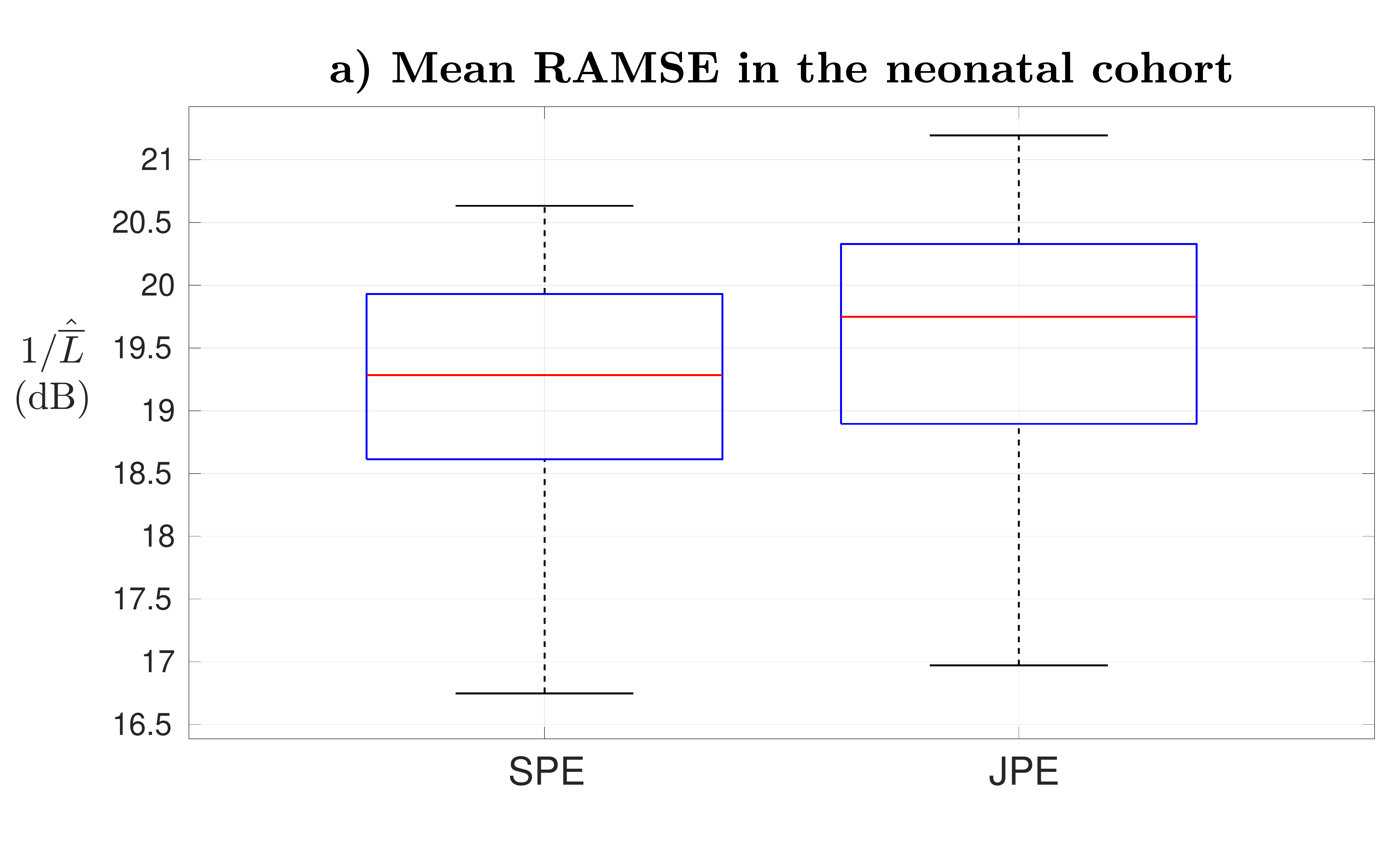}
\end{minipage}
\begin{minipage}{0.49\textwidth}
\begin{center}
\abovebaseline[0pt]{\includegraphics[width=0.28\textwidth]{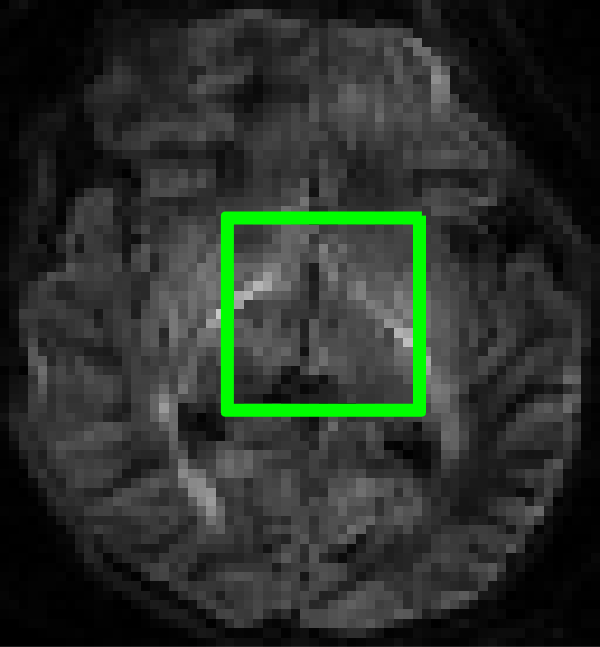}}
\abovebaseline[0pt]{\includegraphics[width=0.262\textwidth]{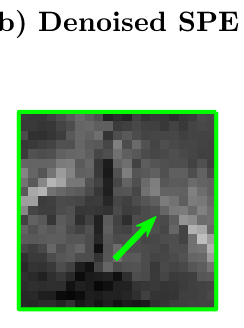}}
\abovebaseline[0pt]{\includegraphics[width=0.418\textwidth]{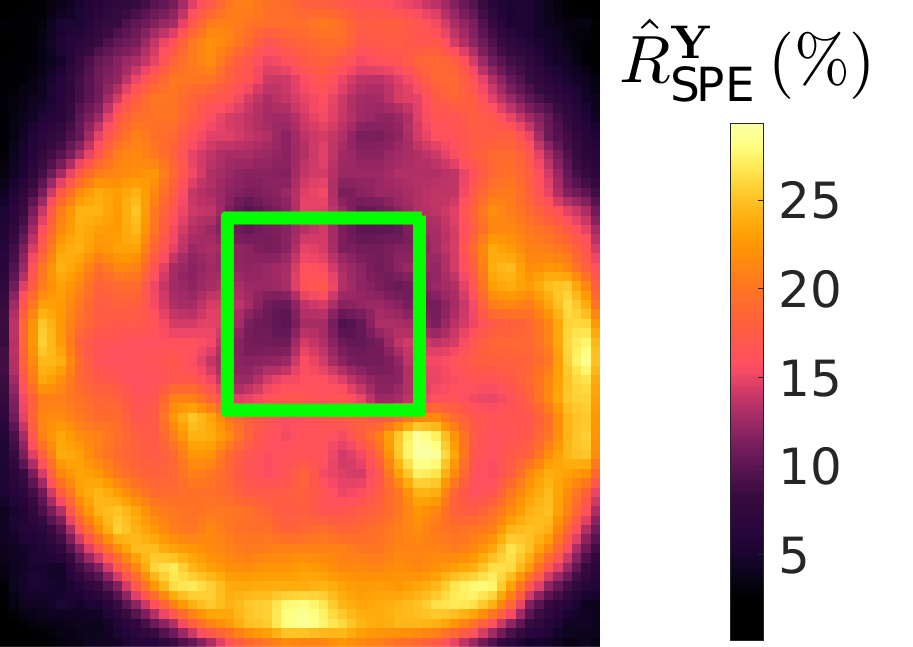}}\\
\abovebaseline[0pt]{\includegraphics[width=0.28\textwidth]{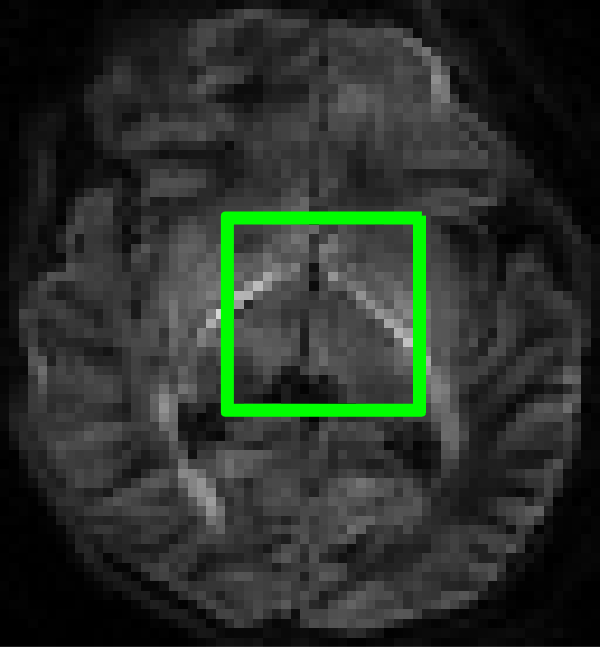}}
\abovebaseline[0pt]{\includegraphics[width=0.262\textwidth]{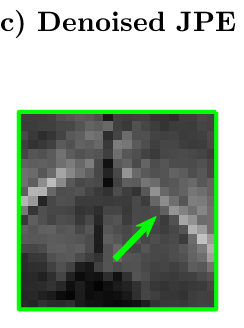}}
\abovebaseline[0pt]{\includegraphics[width=0.418\textwidth]{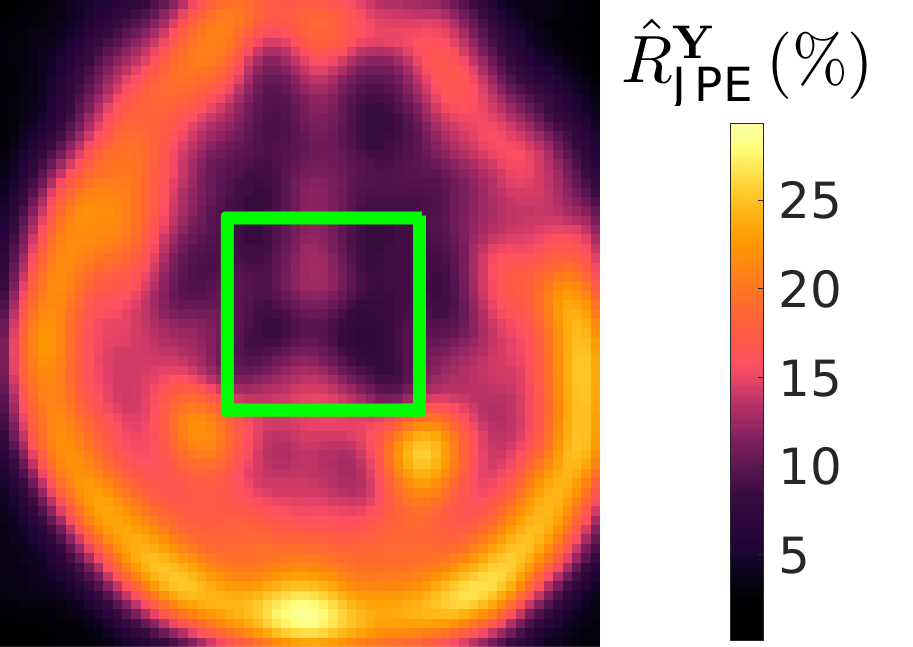}}
\end{center}
\end{minipage}
\caption{\textbf{a)} Distributions of the estimated RAMSE for denoising using the volumes of each PE separately (SPE) and all volumes together (JPE). \textbf{b,c)} From left to right, denoising for \textbf{b)} SPE ($\hat{\gamma}=0.95$ for all PEs) and \textbf{c)} JPE ($\hat{\gamma}=0.8$) for an exemplary slice at $b=2.6\text{ms}/\mu\text{m}^2$; zoomed results; rank estimates $\hat{R}$, given as a percentage of the number of volumes $N$.}
\label{fig:FOPQ}
\end{figure}

In the neonatal case, four PE directions were interleaved throughout the acquisition, each of them generating its own spatial noise distribution. Two strategies to deal with these datasets are compared. First, the problem is reduced to four independent denoising subproblems that use the volumes corresponding to each PE separately (SPE method). Second, the noise model in~\eqref{ec:YMIN} is considered so the different PEs are treated jointly (JPE method). The comparisons are performed at a group level using a sample of $87$ subjects chosen uniformly at random from the set of complete scans within the neonatal cohort and using $\mathbf{G}=\mathbf{G}_{\text{RA}}$ for both strategies. The inverse RAMSE averaged on automatically extracted brain voxels is compared in Fig.~\ref{fig:FOPQ}a. A reduction of the average RAMSE $\hat{\overline{L}}$ has been observed when using the JPE approach with an average improvement ratio of $\rho=0.39\,\mbox{dB}$. These results give a negligible $p$-value when performing a paired right-tailed sign test against the null hypothesis that the median of $\hat{\overline{L}}_{\text{JPE}}-\hat{\overline{L}}_{\text{SPE}}$ is greater than zero or zero. Left column of Figs.~\ref{fig:FOPQ}b,c shows that the recovered brain signal appears similar at coarse scales. However, when zooming in, as in the central column, better delineated edges are observed when using the JPE model, as observed in the ascending tract feature highlighted by the arrow. The right column explains this effect by resorting to the rank estimates, with JPE estimated signal ranks generally lower and smoother than the sum of SPE estimates for each PE; more data samples have been made available by leveraging all PEs so that any residual redundancy in the information contained in the different PEs will increment the denoising potential.

\subsection{Complex denoising for unbiased diffusion measures}

\label{sec:EDDM}

\begin{figure}[!htb]
\begin{center}
\includegraphics[width=\textwidth]{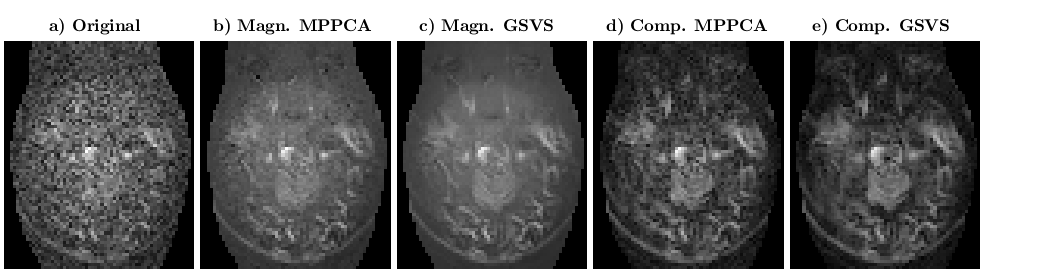}
\includegraphics[width=\textwidth]{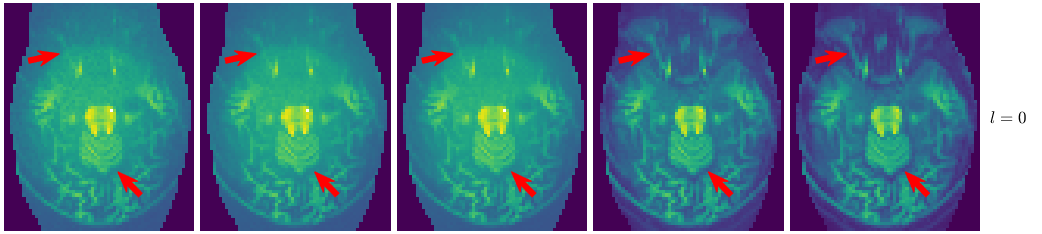}
\includegraphics[width=\textwidth]{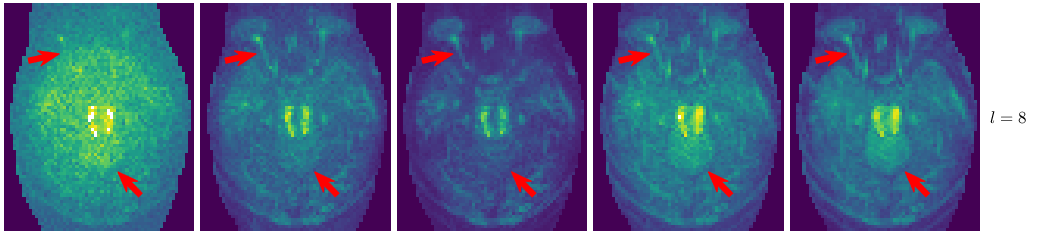}
\end{center}
\caption{\textbf{a)} Original dataset. \textbf{b,c)} Magnitude data denoising by \textbf{b)} MPPCA; \textbf{c)} adapting GSVS. \textbf{d,e)} Complex data denoising by \textbf{d)} adapting MPPCA; \textbf{e)} GSVS. Top: denoising results in an exemplary slice and orientation at $b=10\text{ms}/\mu\text{m}^2$. Mid and bottom: estimates of the spherical harmonic power for the same shell and slice respectively for orders $l=0$ and $l=8$. All results correspond to the optimal aspect ratio $\hat{\gamma}=0.85$.}
\label{fig:ERAD}
\end{figure}

This Section provides visual evidence of the benefits of operating with complex data for denoising and performs comparisons with the MPPCA method in~\cite{Veraart16b}, which uses the EXP2 noise estimator in~\eqref{ec:NET1}. To separately assess the technical refinements introduced in matrix recovery, comparisons are also performed when adapting this method to operate on complex data. Due to complexity of neonatal and fetal data preprocessing for motion and distortion correction and the limited applicability of MPPCA under general noise models, we use diffusion measures of non-preprocessed adult data. Similar properties of magnitude and complex data retrieval have been observed in neonatal and fetal examples. Fig.~\ref{fig:ERAD} shows the denoising results for the $b=10\text{ms}/\mu\text{m}^2$ shell using an exemplary orientation (top) together with the estimated spherical harmonic power obtained with \textsmaller{\textsc{MRtrix3}}~\citep{Tournier19} at orders $l=0$ (mid) and $l=8$ (bottom). We illustrate the results on an inferior slice with particularly low SNR and affected by strong non-linear phase corruption. The denoising capabilities are evident when comparing with the original data in Fig.~\ref{fig:ERAD}a. However, the retrieved signal differs noticeably for the tested approaches. Magnitude-only denoising in Figs.~\ref{fig:ERAD}b, using MPPCA, and~\ref{fig:ERAD}c, adapting our GSVS method (assuming the same noise model but in a real instead of a complex domain), provides biased estimates when compared to complex denoising in Figs.~\ref{fig:ERAD}d, adapting MPPCA, and~\ref{fig:ERAD}e, using GSVS, especially in free diffusion areas. The bias reduces the contrast between restricted and free diffusion, with hampered detectability of the optical nerve for $l=0$ (highlighted by the arrows). Complex denoising also appears more effective in preserving the diffusion features for $l=8$, for instance at the cerebellum (highlighted by the arrows). Differences between data-based noise estimation and hard thresholding (Fig.~\ref{fig:ERAD}d) versus propagation of noise measures and optimal shrinkage (Fig.~\ref{fig:ERAD}e) are more subtle. However, it appears that the high spatial frequency harmonics are more accurately recovered when using the methods here proposed, with more compactly structured features in~\ref{fig:ERAD}e versus~\ref{fig:ERAD}d (top row and $l=8$). This is supported by the random matrix error estimates, which predict an average AMSE reduction of $1.84\,\mbox{dB}$ (ranging from $1.10$ to $5.12\,\mbox{dB}$ at different locations within the FOV) when using our method versus the complex version of~\cite{Veraart16b}.

\section{Discussion}

\label{sec:DISC}

We have presented a signal recovery algorithm for DWI based on singular value shrinkage of Casorati matrices with local spatial information and diffusion measures respectively arranged along the rows and the columns. Due to the limited diversity of spatial and diffusion information, particularly when correcting for unpredictable phase variations from bulk motion during the application of the diffusion gradients, it is reasonable to assume that the signal component within these matrices is approximately low rank, while the sampling noise properties can be accurately modeled when the scanner noise measures are propagated through the reconstruction. This allows the application of recent results on the asymptotic limits of additive perturbations of random matrices, most essentially the confinement of the noise bulk singular values, to obtain an objective estimate of the signal components. The method prevents recovery biases by using complex data information and has been applied to different brain DWI cohorts with scans obtained by advanced protocols that make use of SMS accelerated and interleaved encodings. Quantitative experiments have been conducted to validate the different assumptions and to compare to a related approach taking advantage of the estimates of the asymptotic error as provided by the theory. Our proposal guarantees optimal asymptotic risks when estimating diffusion information by singular value mappings, which stems from a synergistic combination of recent results on signal prediction using random matrix theory and the usage of the scanner noise measures.

Phase retrieval and correction has been proposed as a separate step to obtain a diffusion measure with additive noise properties, for instance, in~\cite{Prah10,Eichner15}. These methodologies, however, inherit the limitations of the implemented phase recovery strategy, namely, potential resolution loss when low pass filtering~\citep{Prah10}, or non comprehensive use of available information when operating slicewise~\citep{Eichner15}. In contrast, our approach considers all the available information to simultaneously estimate the magnitude and the phase of the signal, so the retrieved signal accuracy is boosted by any given amount of redundancy in the complex data. Thus, although the variance of the estimates could potentially be reduced with localized phase correction, we have resorted to a simple but robust global linear correction. Our rationale is that, when compared to local phase manipulations, this correction has reduced risks of altering the statistical properties of the data and consequently biasing the complex denoising results, particularly when considering very low SNR regimes.

Patch construction in our denoising method is based on spatial proximity. However, the estimates could benefit from more refined locally adaptive criteria to build the spatial patches that also take into account the signal similarities~\citep{Katkovnik10} as well as from patches being employed both in the spatial and the diffusion coordinates. Recent results on singular value shrinkage denoising~\citep{Leeb19} show that improved estimates are possible when either the noise or the signal cannot be assumed unitarily invariant by using weighted Frobenius losses respectively after noise standardization or after an appropriate decomposition into a set of subproblems operating on subspaces of the original problem. This idea of obtaining the estimates by superposing different singular value shrinkage subproblems is actually connected with our heuristic usage of superposed patches in the image domain, so our current efforts are focused on methods to inform patch construction or problem decomposition using the theory in~\cite{Leeb19}.

Finally, attention should be paid to other degrading factors that may increase the rank of the matrices considered, most notably motion and non stationary distortions. In this regard, similarly to the phase correction adopted in this paper, the denoising model could include other correction steps prone to promote low-rankness while keeping the noise propagation tractability. In case that the low rank assumptions or the asymptotic regime become less justifiable, the matrix recovery problem may be reformulated using different estimation criteria~\citep{Yadav17}.

\section{Conclusions}

\label{sec:CONC}

We have proposed a method for patch-based DWI retrieval based on random matrix theory. Our proposal extends previous contributions in the field by pushing the application of the theory to discontinuous, correlated and temporally heteroscedastic noise and considering more refined manipulations on the empirical singular values. Importantly, it overcomes the limitations inherent to the simultaneous estimation of the signal and the noise properties; in our approach, the latter are modelled by the propagation of the scanner measures through the reconstruction and spectral decomposition operators. Moreover, the objective nature of this approach to denoising, using only the asymptotic properties of additive low rank perturbations of noisy matrices to find the optimal denoising strength without resorting to empirical parameter tuning, is used for patch size selection by means of the RAMSE estimates. Experiments have been conducted quantifying the benefits of noise modeling and SV shrinkage versus noise estimation and SV truncation as previously proposed, with simulations showing that our method compares favourably with related and alternative state of the art approaches. In addition, evidence has been provided on improvements when promoting low-rankness by phase demodulation and on efficient operation in accelerated scanning regimes including in-plane spectral subsampling, SMS and PF as well as in temporally interleaved acquisition strategies. Finally, we have demonstrated the ability to produce denoised and debiased DWI estimates in challenging fetal, neonatal and adult neuroimaging applications.

\section*{Acknowledgments}



This work received funding from the European Research Council under the European Union's Seventh Framework Programme (FP7/20072013/ERC grant agreement no. [319456] dHCP project). The research was supported by the Wellcome/EPSRC Centre for Medical Engineering at King's College London [WT 203148/Z/16/Z]; the Medical Research Council [MR/K006355/1]; and the National Institute for Health Research (NIHR) Biomedical Research Centre based at Guy's and St Thomas' NHS Foundation Trust and King's College London. The views expressed are those of the authors and not necessarily those of the NHS, the NIHR or the Department of Health. The authors also acknowledge the Department of Perinatal Imaging \& Health at King's College London.

\clearpage

\bibliographystyle{apalike}
\bibliography{ComplexMP}

\clearpage


\end{document}